\documentclass[journal]{IEEEtran}
\usepackage{cite}

\ifCLASSINFOpdf
   \usepackage[pdftex]{graphicx}
  \graphicspath{{./figs/}}
  \DeclareGraphicsExtensions{.pdf,.png,.jpeg,.jpg}
\else
\fi
\usepackage{array}

  \usepackage[caption=false,font=footnotesize]{subfig}

\usepackage{stfloats}
\usepackage{url}

\usepackage[T1]{fontenc}
\usepackage[utf8]{inputenc}
\usepackage{siunitx}
\DeclareSIUnit{\sample}{S}
\DeclareSIUnit{\belmilliwatt}{Bm}
\DeclareSIUnit{\dBm}{\deci\belmilliwatt}
\usepackage{xcolor}
\usepackage{acro}

\usepackage{multirow} 
\usepackage{makecell} 

\usepackage{mathtools}
\usepackage{amssymb,amsfonts}

\newcommand{\ie}{i.\,e.\ }		

\usepackage[scr=dutchcal,scrscaled=1.06]{mathalfa}	
\usepackage{arydshln} 
\usepackage{bm}
\usepackage{upgreek}

\DeclareRobustCommand{\mathbfup}[1]{\begingroup\changegreekbf\mathbf{#1}\endgroup}

\makeatletter
\def\changegreek{\@for\next:={%
		alpha,beta,gamma,delta,epsilon,zeta,eta,theta,kappa,lambda,mu,nu,xi,pi,rho,sigma,%
		tau,upsilon,phi,chi,psi,omega,varepsilon,vartheta,varpi,varrho,varsigma,varphi}%
	\do{\expandafter\let\csname\next\expandafter\endcsname\csname up\next\endcsname}}
\def\changegreekbf{\@for\next:={%
		alpha,beta,gamma,delta,epsilon,zeta,eta,theta,kappa,lambda,mu,nu,xi,pi,rho,sigma,%
		tau,upsilon,phi,chi,psi,omega,varepsilon,vartheta,varpi,varrho,varsigma,varphi}%
	\do{\expandafter\def\csname\next\expandafter\endcsname\expandafter{%
			\expandafter\bm\expandafter{\csname up\next\endcsname}}}}
\makeatother

\makeatletter
\def\iddots{\mathinner{\mkern1mu\raise\p@
		\hbox{.}\mkern2mu\raise4\p@\hbox{.}\mkern2mu
		\raise7\p@\vbox{\kern7\p@\hbox{.}}\mkern1mu}}
\makeatother

\newcommand*{\figref}[1]{Fig.~\ref{#1}}
\newcommand*{\tabref}[1]{Table~\ref{#1}}
\newcommand*{\secref}[1]{Section~\ref{#1}}
\newcommand*{\imag}{\ensuremath{j}}
\newcommand*{\abs}[1]{\left|#1\right|}
\newcommand*{\norm}[1]{\left\lVert#1\right\rVert}
\newcommand*{\floor}[1]{\left\lfloor#1\right\rfloor}

\newcommand*{\vect}[1]{\ensuremath{\mathbfup{#1}}}
\newcommand*{\matx}[1]{\ensuremath{\mathbfup{#1}}}
\newcommand*{\tran}{^{\ensuremath{\mathsf{T}}}}
\newcommand*{\herm}{^{\ensuremath{\mathsf{H}}}}
\DeclareMathOperator{\expv}{\mathrm{E}}

\DeclareMathOperator{\sign}{sgn}

\renewcommand{\arraystretch}{1.2}

\DeclareAcronym{AAF}{
	short = {AAF},
	short-indefinite = {an},
	long  = {anti-aliasing filter},
	long-indefinite = {an}
}
\DeclareAcronym{ADC}{
	short = {ADC},
	short-indefinite = {an},
	long  = {analog-to-digital converter},
	long-indefinite = {an}
}
\DeclareAcronym{AF}{
	short = {AF},
	short-indefinite = {an},
	long  = {adaptive filter},
	long-indefinite = {an}
}
\DeclareAcronym{AFE}{
	short = {AFE},
	short-indefinite = {an},
	long  = {analog front-end},
	long-indefinite = {an}
}
\DeclareAcronym{ALD}{
	short = {ALD},
	short-indefinite = {an},
	long  = {approximate linear dependency},
	long-indefinite = {an}
}
\DeclareAcronym{ANN}{
	short = {ANN},
	short-indefinite = {an},
	long  = {artificial neural network},
	long-indefinite = {an}
}
\DeclareAcronym{ASIC}{
	short = {ASIC},
	short-indefinite = {an},
	long  = {application-specific integrated circuit},
	long-indefinite = {an}
}
\DeclareAcronym{Aux}{
	short = {Aux},
	short-indefinite = {an},
	long  = {auxiliary},
	long-indefinite = {an}
}

\DeclareAcronym{BB}{
	short = {BB},
	long  = {baseband}
}
\DeclareAcronym{BSS}{
	short = {BSS},
	long  = {blind source separation}
}

\DeclareAcronym{CA}{
	short = {CA},
	long  = {carrier aggregation}
}
\DeclareAcronym{CFO}{
	short = {CFO},
	long = {carrier frequency offset}
}
\DeclareAcronym{CMOS}{
	short = {CMOS},
	long  = {complementary metal–oxide–semiconductor}
}
\DeclareAcronym{CP}{
	short = {CP},
	long = {cyclic prefix}
}
\DeclareAcronym{CR}{
	short = {CR},
	long = {clipping ratio}
}
\DeclareAcronym{CSF}{
	short = {CSF},
	long  = {channel-select filter}
}

\DeclareAcronym{DAC}{
	short = {DAC},
	long  = {digital-to-analog converter}
}
\DeclareAcronym{DC}{
	short = {DC},
	long  = {direct current}
}
\DeclareAcronym{DCT}{
	short = {DCT},
	long  = {discrete cosine transform}
}
\DeclareAcronym{DFE}{
	short = {DFE},
	long  = {digital-front end}
}
\DeclareAcronym{DFT}{
	short = {DFT},
	long  = {discrete Fourier transform}
}
\DeclareAcronym{DIC}{
	short = {DIC},
	long  = {digital interference cancellation}
}
\DeclareAcronym{DL}{
	short = {DL},
	long  = {downlink}
}
\DeclareAcronym{DIM}{
	short = {DSIM},
	long  = {digital interference mitigation}
}
\DeclareAcronym{DSIM}{
	short = {DSIM},
	long  = {digital self-interference mitigation}
}

\DeclareAcronym{FDD}{
	short = {FDD},
	short-indefinite = {an},
	long  = {frequency-division duplex}
}
\DeclareAcronym{FFT}{
	short = {FFT},
	short-indefinite = {an},
	long  = {fast Fourier transform}
}
\DeclareAcronym{FIR}{
	short = {FIR},
	short-indefinite = {an},
	long  = {finite impulse response}
}
\DeclareAcronym{FPGA}{
	short = {FPGA},
	short-indefinite = {an},
	long  = {field-programmable gate array}
}
\DeclareAcronym{FSM}{
	short = {FSM},
	short-indefinite = {an},
	long  = {finite-state machine}
}

\DeclareAcronym{GSM}{
	short = {GSM},
	long  = {Global System for Mobile Communications}
}

\DeclareAcronym{I}{
	short = {I},
	short-indefinite = {an},
	long = {in-phase},
	long-indefinite = {an}
}
\DeclareAcronym{IBFD}{
	short = {IBFD},
	short-indefinite = {an},
	long  = {in-band full-duplex},
	long-indefinite = {an}
}
\DeclareAcronym{ICN}{
	short = {ICN},
	short-indefinite = {an},
	long  = {interference-to-carrier-plus-noise},
	long-indefinite = {an}
}
\DeclareAcronym{IDFT}{
	short = {IDFT},
	short-indefinite = {an},
	long  = {inverse discrete Fourier transform},
	long-indefinite = {an}
}
\DeclareAcronym{IF}{
	short = {IF},
	short-indefinite = {an},
	long  = {intermediate frequency},
	long-indefinite = {an}
}
\DeclareAcronym{IIP2}{
	short = {IIP2},
	short-indefinite = {an},
	long  = {second-order input intercept point}
}
\DeclareAcronym{IIR}{
	short = {IIR},
	short-indefinite = {an},
	long  = {infinite impulse response},
	long-indefinite = {an}
}
\DeclareAcronym{IMD}{
	short = {IMD},
	short-indefinite = {an},
	long  = {intermodulation distortion},
	long-indefinite = {an}
}
\DeclareAcronym{IQ}{
	short = {I/Q},
	short-indefinite = {an},
	long  = {in-phase and quadrature},
	long-indefinite = {an}
}
\DeclareAcronym{IRR}{
	short = {IRR},
	short-indefinite = {an},
	long  = {image rejection ratio},
	long-indefinite = {an}
}

\DeclareAcronym{KAF}{
	short = {KAF},
	long  = {kernel adaptive filter}
}
\DeclareAcronym{KLT}{
	short = {KLT},
	long  = {Karhunen-Loève transform}
}
\DeclareAcronym{KRLS}{
	short = {KRLS},
	long  = {kernel recursive least squares}
}

\DeclareAcronym{LMMSE}{
	short = {LMMSE},
	short-indefinite = {an},
	long  = {linear minimum mean square error}
}
\DeclareAcronym{LMS}{
	short = {LMS},
	short-indefinite = {an},
	long  = {least mean squares}
}
\DeclareAcronym{LNA}{
	short = {LNA},
	short-indefinite = {an},
	long  = {low-noise amplifier}
}
\DeclareAcronym{LO}{
	short = {LO},
	short-indefinite = {an},
	long  = {local oscillator}
}
\DeclareAcronym{LS}{
	short = {LS},
	short-indefinite = {an},
	long  = {least squares}
}
\DeclareAcronym{LTE}{
	short = {LTE},
	short-indefinite = {an},
	long  = {Long-Term Evolution}
}

\DeclareAcronym{MB}{
	short = {MB},
	short-indefinite = {an},
	long  = {model-based}
}
\DeclareAcronym{MB-LMS}{
	short = {MB-LMS},
	short-indefinite = {an},
	long  = {model-based LMS}
}
\DeclareAcronym{MSE}{
	short = {MSE},
	short-indefinite = {an},
	long  = {mean square error}
}
\DeclareAcronym{MSIC}{
	short = {MSIC},
	short-indefinite = {an},
	long  = {mixed-signal interference cancellation}
}

\DeclareAcronym{N-LMS}{
	short = {N-LMS},
	short-indefinite = {an},
	long  = {normalized LMS}
}
\DeclareAcronym{NMSE}{
	short = {NMSE},
	short-indefinite = {an},
	long  = {normalized mean square error}
}
\DeclareAcronym{NR}{
	short = {NR},
	short-indefinite = {an},
	long  = {New Radio}
}

\DeclareAcronym{OFDM}{
	short = {OFDM},
	short-indefinite = {an},
	long = {orthogonal frequency-division multiplexing},
	long-indefinite = {an}
}
\DeclareAcronym{OFDMA}{
	short = {OFDMA},
	short-indefinite = {an},
	long = {orthogonal frequency-division multiple access},
	long-indefinite = {an}
}

\DeclareAcronym{PA}{
	short = {PA},
	long  = {power amplifier}
}
\DeclareAcronym{PCB}{
	short = {PCB},
	long  = {printed circuit board}
}
\DeclareAcronym{PDF}{
	short = {PDF},
	long  = {probability density function}
}
\DeclareAcronym{PSD}{
	short = {PSD},
	long  = {power spectral density}
}

\DeclareAcronym{Q}{
	short = {Q},
	long = {quadrature}
}

\DeclareAcronym{RB}{
	short = {RB},
	short-indefinite = {an},
	long  = {resource block}
}
\DeclareAcronym{RF}{
	short = {RF},
	short-indefinite = {an},
	long  = {radio frequency}
}
\DeclareAcronym{RLS}{
	short = {RLS},
	short-indefinite = {an},
	long  = {recursive least squares}
}
\DeclareAcronym{RMS}{
	short = {RMS},
	short-indefinite = {an},
	long  = {root mean square}
}
\DeclareAcronym{RV}{
	short = {RV},
	short-indefinite = {an},
	long = {random variable}
}
\DeclareAcronym{Rx}{
	short = {Rx},
	short-indefinite = {an},
	long  = {receive}
}

\DeclareAcronym{SAF}{
	short = {SAF},
	long  = {spline adaptive filter}
}
\DeclareAcronym{SCFDMA}{
	short = {SC-FDMA},
	short-indefinite = {an},
	long = {single-carrier frequency-division multiple access}
}
\DeclareAcronym{SCT}{
	short = {SCT},
	short-indefinite = {an},
	long = {sliding cosine transform}
}
\DeclareAcronym{SGD}{
	short = {SGD},
	short-indefinite = {an},
	long  = {stochastic gradient descent}
}
\DeclareAcronym{SIC}{
	short = {SIC},
	long  = {self-interference cancellation}
}
\DeclareAcronym{SINR}{
	short = {SINR},
	short-indefinite = {an},
	long  = {signal-to-interference-plus-noise ratio}
}
\DeclareAcronym{SNR}{
	short = {SNR},
	short-indefinite = {an},
	long  = {signal-to-noise ratio}
}
\DeclareAcronym{SVM}{
	short = {SVM},
	short-indefinite = {an},
	long  = {support-vector machine}
}

\DeclareAcronym{TD}{
	short = {TD},
	long  = {transform-domain}
}
\DeclareAcronym{TD-LMS}{
	short = {TD-LMS},
	long  = {transform-domain LMS}
}
\DeclareAcronym{TDD}{
	short = {TDD},
	long  = {time-division duplex}
}
\DeclareAcronym{TDE}{
	short = {TDE},
	long  = {time-delay estimation}
}
\DeclareAcronym{Tx}{
	short = {Tx},
	long  = {transmit}
}

\DeclareAcronym{UL}{
	short = {UL},
	short-indefinite = {an},
	long  = {uplink},
	long-indefinite = {an}
}

\DeclareAcronym{VSA}{
	short = {VSA},
	long = {vector signal analyzer}
}
\DeclareAcronym{VSG}{
	short = {VSG},
	long = {vector signal generator}
}

\DeclareAcronym{WGN}{
	short = {WGN},
	long = {white Gaussian noise}
}
\DeclareAcronym{WSS}{
	short = {WSS},
	long  = {wide-sense stationary}
}

\begin{document}
\title{Spline-Based Adaptive Cancellation of Even-Order Intermodulation Distortions in LTE-A/5G RF Transceivers}

\author{Thomas~Paireder,~\IEEEmembership{Graduate~Student~Member,~IEEE},
		    Christian~Motz,~\IEEEmembership{Graduate~Student~Member,~IEEE},
        and~Mario~Huemer,~\IEEEmembership{Senior~Member,~IEEE}
\thanks{The financial support by the Austrian Federal Ministry for Digital and Economic Affairs, the National Foundation for Research, Technology and Development and the Christian Doppler Research Association is gratefully acknowledged.}%
\thanks{The authors are with the Christian Doppler Laboratory for Digitally Assisted RF Transceivers for Future Mobile Communications, Institute of Signal Processing, Johannes Kepler University Linz, 4040 Linz, Austria (e-mail: thomas.paireder@jku.at).}%
}

\markboth{}%
{}

\maketitle

\begin{abstract}
Radio frequency transceivers operating in in-band full-duplex or frequency-division duplex mode experience strong transmitter leakage. Combined with receiver nonlinearities, this causes intermodulation products in the baseband, possibly with higher power than the desired receive signal. In order to restore the receiver signal-to-noise ratio in such scenarios, we propose two novel digital self-interference cancellation approaches based on spline interpolation. Both employ a Wiener structure, thereby matching the baseband model of the intermodulation effect. Unlike most state-of-the-art spline-based adaptive learning schemes, the proposed concept allows for complex-valued in- and output signals. The optimization of the model parameters is based on the stochastic gradient descent concept, where the convergence is supported by an appropriate step-size normalization. Additionally, we provide a gain control scheme and enable pipelining in order to facilitate a hardware implementation. An optional input transform improves the performance consistency for correlated sequences. In a realistic interference scenario, the proposed algorithms clearly outperform a state-of-the-art least mean squares variant with comparable complexity, which is specifically tailored to second-order intermodulation distortions. The high flexibility of the spline interpolation allows the spline-based Wiener models to match the performance of the kernel recursive least squares algorithm at less than 0.6\,\% of the arithmetic operations.
\end{abstract}

\begin{IEEEkeywords}
	Adaptive learning, intermodulation distortion, LTE, self-interference cancellation, spline interpolation, RF transceivers.
\end{IEEEkeywords}

\IEEEpeerreviewmaketitle

\section{Introduction}
\IEEEPARstart{P}{ower} efficiency is a key aspect for \ac{RF} transceivers in mobile communications equipment in order to ensure sufficient battery life at high data rates. Combined with other metrics, such as cost or area usage, this might lead to a decrease in receiver linearity. Given the low powers of the wanted \ac{Rx} signal at the input, these design trade-offs usually do not cause relevant distortions. However, in \ac{IBFD} and \ac{FDD} operation, a strong \ac{Tx} leakage in the receiver is unavoidable, leading to a severe deterioration of the \ac{Rx} signal. When using the predominant direct-conversion receiver architecture, especially even-order \acp{IMD} are a major issue, since they fall directly into the \ac{Rx} \ac{BB} independent of the \ac{Tx} carrier frequency \cite{Razavi1997}.

\acuse{LTE}
In the Long-Term Evolution Advanced (LTE-A) and 5G \ac{NR} standards, an important transmission mode is \ac{FDD}. The resulting separation of the \ac{Tx} and \ac{Rx} carriers enables the usage of band-selection filters, in communication transceivers these are typically duplexers. These components provide a limited suppression of the transmitter-to-receiver leakage of about \SIrange{50}{55}{\dB} \cite{saw_tdk_b8625}. A higher isolation is not feasible due to disadvantages such as higher cost or increased insertion losses. With \ac{Tx} powers of up to \SI{27}{\dBm} at the output of the \ac{PA}, the spectrally shaped leakage still has considerable power at the duplexer \ac{Rx} port, causing \ac{IMD}. Besides design changes in the \ac{AFE}, several \ac{BB} mitigation techniques exist to attenuate the \ac{IMD} interference \cite{Sadjina2020}. In this work, we focus on fully digital self-interference cancellation for \ac{FDD} transceivers, where the interference is replicated based on the known \ac{Tx} data. While increasing the computational burden in the digital \ac{BB} compared to mixed-signal solutions, \ac{DSIM} approaches do not require any changes to the \ac{AFE} and scale well with smaller technology nodes. Most published approaches target second-order intermodulation products, both, with \cite{Frotzscher2009, Kiayani2017, Gebhard2017, Gebhard2019_J1} and without \cite{Frotzscher2008, Lederer2011} a frequency-selective leakage path. However, also higher-order products are likely to occur but are rarely covered in literature \cite{Keehr2011, Paireder2021}. One category of suitable algorithms are general learning schemes, such as \acp{KAF}, \acp{SVM} or \acp{ANN} with constant or adaptive activation functions \cite{Comminiello2018, Auer2020_C1, Auer2020_C2, Vecci1998, Uncini2003, Bohra2020}. Other concepts are adaptive truncated Volterra series and functional link adaptive filters \cite{Comminiello2013, Comminiello2018}, which both utilize limited model knowledge. However, these models result in an unreasonably high number of parameters to be estimated, slowing down adaptation and increasing complexity. The generation of higher-order \ac{IMD} is best represented by Wiener models, but the required nonlinear estimation is often assumed to be unsuitable for real-time implementation on devices with limited computational resources. With the development of \acp{SAF} \cite{Scarpiniti2013, Scarpiniti2014, Scarpiniti2015}, low-complex adaptive Wiener models for a wide range of nonlinearities are available. However, most of them target real-valued functions and, thus, are not applicable to the complex-valued \ac{IMD} cancellation problem. Two exceptions featuring spline interpolation with complex control points are presented in \cite{Campo2018, Campo2020}. Though, both show specialized concepts without normalization, which are not suited for the \ac{IMD} problem.

In this work, we introduce two novel Wiener \acp{SAF}, comprised of a complex-valued linear system, a fixed internal nonlinearity and a real- or complex-valued spline output. The latter is achieved by combining two real-valued spline functions, where one models the real and one the imaginary part of the output. Both proposed \ac{SAF} variants allow for a filtered output, which, for example, can cover a delayed update as it is caused by a pipelined hardware implementation. For the first time, internal clipping is avoided by means of an optimization constraint, that controls the norm of the linear filter weights. Additionally, an alternative limiter for the weight norm with lower complexity is proposed. The \ac{SGD} update used for learning is improved by means of an appropriate novel step-size normalization. Optionally, the learning is augmented with the well-known \ac{TD} concept to improve the performance consistency of the algorithms.

This paper is organized as follows: In \secref{sec:imd_self_intf}, we provide an in-depth analysis of the \ac{IMD} effect, leading to a \ac{BB} model that is essential for all further considerations. For completeness, in \secref{sec:spline_basics}, we give a brief overview of the spline interpolation concept. In \secref{sec:spline_imdx_canc}, we derive normalized \ac{SGD} update equations for the proposed \acp{SAF}, discuss possible optimizations and assess the computational complexity. \secref{sec:perf_simulation} quantifies the \ac{IMD} cancellation performance of our concepts on a real-world scenario.

\section{Self-Interference Due To Intermodulation Distortions}
\label{sec:imd_self_intf}

\begin{figure*}
	\centering
	\includegraphics[width=0.86\textwidth]{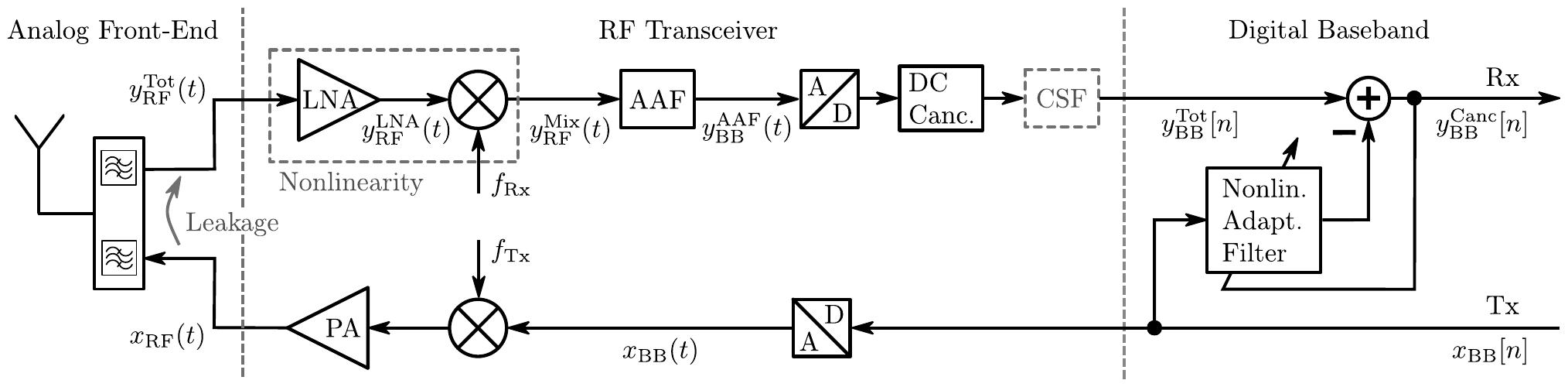}
	\caption{Block diagram of FDD RF transceiver with digital cancellation of IMDx interference.}
	\label{fig:Transceiver}
\end{figure*}%
\figref{fig:Transceiver} schematically depicts one \ac{Tx} and one \ac{Rx} path of \iac{FDD} transceiver, operating simultaneously on a common antenna. As a consequence of non-ideal Tx-Rx isolation in the analog front-end, the \ac{Tx} signal leaks into the receiver, where it causes nonlinear distortions that overlay the wanted \ac{Rx} \ac{BB} signal. Note that in case of carrier aggregation, this problem potentially persists for all combinations of \ac{Tx} and \ac{Rx} chains.

A partial modeling of this effect is shown in \cite{Gebhard2017, Gebhard2019_T1}, which we use as a foundation to derive the interference components that occur in the \ac{Rx} \ac{BB}. Compared to the available literature, we include all important cross-terms of \ac{Tx} leakage, wanted \ac{Rx} and noise. This thorough analysis is essential in order to correctly interpret measurement results on the \ac{IMD} effect. We start with the known digital \ac{Tx} \ac{BB} sequence $x_{\text{BB}}[n]$, which first passes the \ac{DAC}. Next, the analog signal $x_{\text{BB}}(t)$ is up-converted to the carrier $f_{\text{Tx}}$. Since we focus on receiver nonlinearities, for our model we assume the up-conversion mixer and the \ac{PA} to be ideal with a total gain of $A_{\text{PA}}$. This simplification is backed by simulations in \secref{sec:perf_simulation}, where we include the saturation behavior of the \ac{PA}. The cancellation performance of the proposed Wiener \ac{SAF} is not impacted by the \ac{PA} nonlinearity. After up-conversion and amplification, the resulting \ac{Tx} \ac{RF} signal is
\begin{equation}
	x_{\text{RF}}(t) = A_{\text{PA}} \, \Re\left\{x_{\text{BB}}(t) \, e^{\imag 2\pi f_{\text{Tx}} t}\right\}.
	\label{equ:tx_rf_mdl_ideal}
\end{equation}
The leakage path, comprising \ac{RF} switches, diplexers and the duplexer, is modeled as
\begin{equation}
	h_{\text{RF}}^{\text{TxL}}(t) = 2\,\Re\left\{h_{\text{BB}}^{\text{TxL}}(t) \, e^{\imag 2\pi f_{\text{Tx}} t}\right\}
\end{equation}
with the equivalent \ac{BB} impulse response $h_{\text{BB}}^{\text{TxL}}(t)$. Depending on the context, $*$ denotes the time-continuous or time-discrete convolution. The factor $2$ is introduced in the \ac{RF} domain to compensate for the scaling effect of the convolution later on. With these definitions, the \ac{RF} leakage signal can be written as:
\begin{align}
	y_{\text{RF}}^{\text{TxL}} &= x_{\text{RF}}(t) * h_{\text{RF}}^{\text{TxL}}(t)\nonumber\\
	&= \frac{A_{\text{PA}}}{2} \left(x_{\text{BB}}(t) \, e^{\imag 2\pi f_{\text{Tx}} t} + x_{\text{BB}}(t)^* \, e^{-\imag 2\pi f_{\text{Tx}} t}\right)\nonumber\\
	&\hspace{7em} * \left(h_{\text{BB}}^{\text{TxL}}(t) \, e^{\imag 2\pi f_{\text{Tx}} t} + h_{\text{BB}}^{\text{TxL}}(t)^* \, e^{-\imag 2\pi f_{\text{Tx}} t}\right)\nonumber\\
	&= \frac{A_{\text{PA}}}{2} \left(x_{\text{BB}}(t) * h_{\text{BB}}^{\text{TxL}}(t)\right) e^{\imag 2\pi f_{\text{Tx}} t}\nonumber\\
	&\hspace{7em} + \frac{A_{\text{PA}}}{2} \left(x_{\text{BB}}(t) * h_{\text{BB}}^{\text{TxL}}(t)\right)^* e^{-\imag 2\pi f_{\text{Tx}} t} \nonumber\\
	&= \Re\left\{y_{\text{BB}}^{\text{TxL}}(t) \, e^{\imag 2\pi f_{\text{Tx}} t}\right\}.
\end{align}
${y_{\text{BB}}^{\text{TxL}}(t) = x_{\text{BB}}(t) * \tilde{h}_{\text{BB}}^{\text{TxL}}(t)}$ is the \ac{BB} equivalent leakage signal. To shorten the notation, we include the \ac{Tx} gain in the impulse response of the leakage path ${\tilde{h}_{\text{BB}}^{\text{TxL}}(t) = A_{\text{PA}} h_{\text{BB}}^{\text{TxL}}(t)}$. Due to the leakage, the total signal at the receiver input is
\begin{equation}
	y_{\text{RF}}^{\text{Tot}}(t) = y_{\text{RF}}^{\text{Rx}}(t) + \eta_{\text{RF}}(t) + y_{\text{RF}}^{\text{TxL}}(t).
\end{equation}
$y_{\text{RF}}^{\text{Rx}}(t)$ is the desired receive signal at the carrier $f_{\text{Rx}}$
\begin{equation}
	y_{\text{RF}}^{\text{Rx}}(t) = \Re\left\{y_{\text{BB}}^{\text{Rx}}(t) \, e^{\imag 2\pi f_{\text{Rx}} t}\right\}
\end{equation}
and $\eta_{\text{RF}}(t)$ is additive thermal noise from the antenna. Both components passed the duplexer and are therefore limited to the bandwidth of the selected \ac{LTE}/\ac{NR} band. However, any noise components that are added after the duplexer might have substantially higher bandwidth. In order to limit the complexity of the model, we neglect any wideband noise in the following and define
\begin{equation}
	\eta_{\text{RF}}(t) = \Re\left\{\eta_{\text{BB}}(t) \, e^{\imag 2\pi f_{\text{Rx}} t}\right\}.
\end{equation}
Any components up to the quadrature mixer (I/Q mixer) potentially exhibit nonlinear behavior, which is modeled to be concentrated in the \ac{LNA} and the mixer.

Based on measurements of the \ac{IMD} products generated by an integrated \ac{CMOS} receiver we assume a polynomial nonlinearity of degree 3 for modeling the \ac{LNA} \cite{Ahmed2013, Mollen2018}:
\begin{equation}
	y_{\text{RF}}^{\text{LNA}}(t) = \alpha_1 \, y_{\text{RF}}^{\text{Tot}}(t) + \alpha_2 \left(y_{\text{RF}}^{\text{Tot}}(t)\right)^2 + \alpha_3 \left(y_{\text{RF}}^{\text{Tot}}(t)\right)^3.
\end{equation}
The coefficients $\alpha_i$ are real-valued quantities. Inserting $y_{\text{RF}}^{\text{Tot}}(t)$ into the \ac{LNA} model yields
\begin{align}
	y_{\text{RF}}^{\text{LNA}}(t) &= \alpha_1 \left(y_{\text{RF}}^{\text{Rx}}(t) + \eta_{\text{RF}}(t) + y_{\text{RF}}^{\text{TxL}}(t)\right) \nonumber\\
	&\phantom{{}={}} + \alpha_2 \, \Big(\!\left(y_{\text{RF}}^{\text{Rx}}(t)\right)^2 + \left(\eta_{\text{RF}}(t)\right)^2 + \left(y_{\text{RF}}^{\text{TxL}}(t)\right)^2 \nonumber\\
	&\phantom{{}={} + \alpha_2 \, \Big(} + 2 \, y_{\text{RF}}^{\text{Rx}}(t) \, \eta_{\text{RF}}(t) + 2 \, y_{\text{RF}}^{\text{Rx}}(t) \, y_{\text{RF}}^{\text{TxL}}(t) \nonumber\\
	&\phantom{{}={} + \alpha_2 \, \Big(} + 2 \, \eta_{\text{RF}}(t) \, y_{\text{RF}}^{\text{TxL}}(t)\Big) \nonumber\\
	&\phantom{{}={}} + \alpha_3 \, \Big(\!\left(y_{\text{RF}}^{\text{Rx}}(t)\right)^3 + \left(\eta_{\text{RF}}(t)\right)^3 + \left(y_{\text{RF}}^{\text{TxL}}(t)\right)^3 \nonumber\\
	&\phantom{{}={} + \alpha_3 \, \Big(} + 3 \left(y_{\text{RF}}^{\text{Rx}}(t)\right)^2 \eta_{\text{RF}}(t) + 3 \left(y_{\text{RF}}^{\text{Rx}}(t)\right)^2 y_{\text{RF}}^{\text{TxL}}(t) \nonumber\\
	&\phantom{{}={} + \alpha_3 \, \Big(} + 3 \, y_{\text{RF}}^{\text{Rx}}(t) \, \left(\eta_{\text{RF}}(t)\right)^2 + 3 \, y_{\text{RF}}^{\text{Rx}}(t) \left(y_{\text{RF}}^{\text{TxL}}(t)\right)^2 \nonumber\\
	&\phantom{{}={} + \alpha_3 \, \Big(} + 3 \left(\eta_{\text{RF}}(t)\right)^2 y_{\text{RF}}^{\text{TxL}}(t) + 3 \, \eta_{\text{RF}}(t) \, \left(y_{\text{RF}}^{\text{TxL}}(t)\right)^2 \nonumber\\
	&\phantom{{}={} + \alpha_3 \, \Big(} + 6 \, y_{\text{RF}}^{\text{Rx}}(t) \, \eta_{\text{RF}}(t) \, y_{\text{RF}}^{\text{TxL}}(t) \Big).
\end{align}
Due to the large number of terms, we refrain from inserting the equivalent \ac{BB} definitions of $y_{\text{RF}}^{\text{Rx}}(t)$, $\eta_{\text{RF}}(t)$ and $y_{\text{RF}}^{\text{TxL}}(t)$.
\begin{table}[!t]
	\renewcommand{\arraystretch}{1.5}
	\caption{Carrier frequencies of signal components after LNA nonlinearity.}
	\label{tab:carrier_freq_lna_outp}
	\centering
	$\begin{array}{l|c}
		\hline
			\text{Summands} & \text{Contained Carriers}\\
		\hline
			y_{\text{RF}}^{\text{Rx}}, \eta_{\text{RF}} & \pm f_{\text{Rx}}\\
			y_{\text{RF}}^{\text{TxL}} & \pm f_{\text{Tx}}\\
			(y_{\text{RF}}^{\text{Rx}})^2, (\eta_{\text{RF}})^2, y_{\text{RF}}^{\text{Rx}} \, \eta_{\text{RF}} & 0, \pm 2 f_{\text{Rx}}\\
			(y_{\text{RF}}^{\text{TxL}})^2 & 0, \pm 2 f_{\text{Tx}}\\
			y_{\text{RF}}^{\text{Rx}} \, y_{\text{RF}}^{\text{TxL}}, \eta_{\text{RF}} \, y_{\text{RF}}^{\text{TxL}} & \pm f_{\text{Rx}} \pm f_{\text{Tx}}\\
			(y_{\text{RF}}^{\text{Rx}})^3, (\eta_{\text{RF}})^3, (y_{\text{RF}}^{\text{Rx}})^2 \, \eta_{\text{RF}}, y_{\text{RF}}^{\text{Rx}} \, (\eta_{\text{RF}})^2 & \pm f_{\text{Rx}}, \pm 3 f_{\text{Rx}}\\
			(y_{\text{RF}}^{\text{TxL}})^3 & \pm f_{\text{Tx}}, \pm 3 f_{\text{Tx}}\\
			(y_{\text{RF}}^{\text{Rx}})^2 \, y_{\text{RF}}^{\text{TxL}}, (\eta_{\text{RF}})^2 \, y_{\text{RF}}^{\text{TxL}}, y_{\text{RF}}^{\text{Rx}} \, \eta_{\text{RF}} \, y_{\text{RF}}^{\text{TxL}} & \pm f_{\text{Tx}}, \pm 2 f_{\text{Rx}} \pm f_{\text{Tx}}\\
			y_{\text{RF}}^{\text{Rx}} \, (y_{\text{RF}}^{\text{TxL}})^2, \eta_{\text{RF}} \, (y_{\text{RF}}^{\text{TxL}})^2  & \pm f_{\text{Rx}}, \pm f_{\text{Rx}} \pm 2 f_{\text{Tx}}\\
		\hline
	\end{array}$
\end{table}%
Instead, in \tabref{tab:carrier_freq_lna_outp}, we provide an overview of the spectral contents of all terms. Due to space constraints, we dropped the time indices of the signals. While the \ac{LNA} model does not include an explicit \ac{DC} component, signal components around \ac{DC} occur due to intermodulation.

Following the \ac{LNA}, the I/Q mixer performs a direct down-conversion to the complex \ac{BB} using the \ac{LO} frequency $f_{\text{Rx}}$. The model of the mixer covers a \ac{DC} feed-through, the desired down-conversion and an RF-to-LO terminal coupling, quantified by the coefficients $\beta_0$, $\beta_1$ and $\beta_2$, respectively. Hence, we have the following model at the mixer output:
\begin{equation}
	y_{\text{RF}}^{\text{Mix}}(t) = \beta_0 \, y_{\text{RF}}^{\text{LNA}}(t) + \beta_1 \, y_{\text{RF}}^{\text{LNA}}(t) \, e^{-\imag 2\pi f_{\text{Rx}} t} + \beta_2 \left(y_{\text{RF}}^{\text{LNA}}(t)\right)^2.
\end{equation}
$\beta_0$ and $\beta_2$ are possibly complex values, since the mixer is implemented using two independent branches. $\beta_1$ is assumed to be real-valued, which reflects a balanced down-conversion of the I and Q components of the wanted signal. This could be ensured by design or calibration. Besides the down-conversion of the wanted \ac{BB} signal and several distortion terms at \ac{BB}, the mixer also outputs spectral components at multiples of $f_{\text{Tx}}$ and $f_{\text{Rx}}$. We assume that all components at additive combinations of these frequencies are suppressed by the \ac{AAF}. A high \ac{Tx} power and thus, strong leakage usually coincides with a low \ac{Rx} power \cite{Motz2020_C1}. Therefore, we do not list all individual, usually weak intermodulation products of $y_{\text{BB}}^{\text{Rx}}(t)$ and $\eta_{\text{BB}}(t)$ but consider them as additive noise $\eta_{\text{BB}}^{\text{IMD}}(t)$.

Components at subtractive combinations of the carriers, \ie located at the offsets ${f_{\Delta} = \pm (a f_{\text{Tx}} - b f_{\text{Rx}})}$, may fall within the \ac{AAF} bandwidth. In case of ${\abs{f_{\Delta}} \leq 6 \, \text{BW}_{\text{Rx}}}$ they potentially overlap with the main \ac{IMD} components centered at \ac{DC}. Here, $a$, $b$ are positive integers and $\text{BW}_{\text{Rx}}$ is the \ac{Rx} bandwidth. The factor $6$ is related to the combined degree of the nonlinearities of the \ac{LNA} and the mixer. We assume that the bandwidth of $y_{\text{BB}}^{\text{TxL}}(t)$ is less than $\text{BW}_{\text{Rx}}$. One example is given by ${f_{\text{Rx}} \approx 2 f_{\text{Tx}}}$, which can occur in case of \ac{CA}. Unlike the \ac{IMD} components centered at \ac{DC}, the intermodulation products around $f_{\Delta}$ are of even and odd order. The most significant components are of the form ${c_k \! \left(y_{\text{BB}}^{\text{TxL}}(t)^*\right)^p \! \left(y_{\text{BB}}^{\text{TxL}}(t)\right)^q} e^{\imag 2\pi f_{\Delta} t}$ with ${p + q \leq 6}$. However, this interference class is not covered in this work. As a result, the subtractive combinations of the carriers are not considered in the estimation process but add to the total noise floor.

At the output of the \ac{AAF}, our assumptions lead to
\begin{align}
	y_{\text{BB}}^{\text{AAF}} &= \frac{1}{2} \alpha_1 \beta_1 y_{\text{BB}}^{\text{Rx}} + \frac{1}{2} \alpha_1 \beta_1 \eta_{\text{BB}} + \eta_{\text{BB}}^{\text{IMD}} \nonumber\\
	&\phantom{{}={}} + \left(\frac{1}{2} \alpha_1^2 \beta_2 + \frac{1}{2} \alpha_2 \beta_0\right) \abs{y_{\text{BB}}^{\text{TxL}}}^2 \nonumber\\
	&\phantom{{}={}} + \left(\frac{3}{4} \alpha_1 \alpha_3 \beta_2 + \frac{3}{8} \alpha_2^2 \beta_2\right) \abs{y_{\text{BB}}^{\text{TxL}}}^4 \nonumber\\
	&\phantom{{}={}} + \frac{5}{16} \alpha_3^2 \beta_2 \abs{y_{\text{BB}}^{\text{TxL}}}^6 \nonumber\\
	&\phantom{{}={}} + \frac{3}{4} \alpha_3 \beta_1 \left(y_{\text{BB}}^{\text{Rx}} + \eta_{\text{BB}}\right) \abs{y_{\text{BB}}^{\text{TxL}}}^2\nonumber\\
	&\phantom{{}={}} + \left(3 \alpha_1 \alpha_3 \beta_2 + \frac{3}{2} \alpha_2^2 \beta_2\right) \left(\abs{y_{\text{BB}}^{\text{Rx}}}^2 + \abs{\eta_{\text{BB}}}^2\right) \abs{y_{\text{BB}}^{\text{TxL}}}^2\nonumber\\
	&\phantom{{}={}} + \frac{45}{16} \alpha_3^2 \beta_2 \left(\abs{y_{\text{BB}}^{\text{Rx}}}^4 + \abs{\eta_{\text{BB}}}^4\right) \abs{y_{\text{BB}}^{\text{TxL}}}^2\nonumber\\
	&\phantom{{}={}} + \frac{45}{16} \alpha_3^2 \beta_2 \left(\abs{y_{\text{BB}}^{\text{Rx}}}^2 + \abs{\eta_{\text{BB}}}^2\right) \abs{y_{\text{BB}}^{\text{TxL}}}^4,
	\label{equ:imdx_mdl_aaf_outp}
\end{align}
where we again dropped the time indices. The first term is the down-converted \ac{Rx} signal and the second and third terms are noise caused by various sources. The pure intermodulation products of the leakage signal are represented by the terms 4 to 6. These components are targeted by the cancellation approaches presented later in this work. The remaining terms of $y_{\text{BB}}^{\text{AAF}}(t)$ are intermodulation products of the leakage with the \ac{Rx} signal or noise. At very high transmit powers, these products could deteriorate the receiver \ac{SINR}. However, due to their dependence on both, $y_{\text{BB}}^{\text{Rx}}(t)$ and $y_{\text{BB}}^{\text{TxL}}(t)$, these terms are difficult to cancel with low to medium hardware complexity. Consequently, we consider them as noise and summarize the signal at the \ac{AAF} output as
\begin{equation}
	y_{\text{BB}}^{\text{AAF}}(t) = A_{\text{lin}} \, y_{\text{BB}}^{\text{Rx}}(t) + \eta_{\text{BB}}^{\text{AAF}}(t) + \sum_{k = 1}^{3} \gamma_k \abs{y_{\text{BB}}^{\text{TxL}}(t)}^{2k}
\end{equation}
with the linear gain ${A_{\text{lin}} = \frac{1}{2} \alpha_1 \beta_1}$, the combined noise $\eta_{\text{BB}}^{\text{AAF}}(t)$ and the combined coefficients $\gamma_k$.

Direct-conversion receivers usually suffer from spurious \ac{DC} components, which, for example, could saturate the \ac{ADC}. This issue is solved by employing a \ac{DC} cancellation stage. Independent of its actual position in the receive chain, we model this stage as a notch filter $h_{\text{DC}}[n]$ in the digital domain directly following the (ideal) \ac{ADC}. Additionally, the digital \ac{BB} signal is commonly limited to the channel bandwidth by the \ac{CSF}. Since this filter might cause issues for the digital \ac{IMD} cancellation, we propose to place the \ac{CSF} after the \ac{DSIM} point. The final digital \ac{BB} model for all following considerations is
\begin{align}
	y_{\text{BB}}^{\text{Tot}}[n] &= \underbrace{A_{\text{lin}} \, y_{\text{BB}}^{\text{Rx}}[n] * h_{\text{DC}}[n]}_{\check{y}_{\text{BB}}^{\text{Rx}}[n]}{} + \check{\eta}_{\text{BB}}[n] \nonumber\\
	&\phantom{{}={}} + \underbrace{\sum_{k = 1}^{3} \gamma_k \abs{x_{\text{BB}}[n] * \tilde{h}_{\text{BB}}^{\text{TxL}}[n]}^{2k} * h_{\text{DC}}[n]}_{y_\text{BB}^{\text{IMD}}[n]}.
	\label{equ:imdx_mdl_bb}
\end{align}
with the filtered \ac{Rx} \ac{BB} signal $\check{y}_{\text{BB}}^{\text{Rx}}[n]$ and the \ac{BB} interference $y_\text{BB}^{\text{IMD}}[n]$. Note that $\check{\eta}_{\text{BB}}[n]$ includes the quantization noise of the \ac{ADC} $\eta_{\text{BB}}^{\text{ADC}}[n]$:
\begin{equation}
	\check{\eta}_{\text{BB}}[n] = \left(\eta_{\text{BB}}^{\text{AAF}}[n] + \eta_{\text{BB}}^{\text{ADC}}[n]\right) * h_{\text{DC}}[n].
\end{equation}
Based on $x_{\text{BB}}[n]$ and $y_{\text{BB}}^{\text{Tot}}[n]$, the \ac{DSIM} algorithm shall estimate the interference and subtract it from the \ac{Rx} \ac{BB} signal. When denoting the replicated interference with $\hat{y}_{\text{BB}}^{\text{IMD}}[n]$, the enhanced \ac{Rx} signal is given by ${y_{\text{BB}}^{\text{Canc}}[n] = y_{\text{BB}}^{\text{Tot}}[n] - \hat{y}_{\text{BB}}^{\text{IMD}}[n]}$.

\section{Basics of Spline Interpolation}
\label{sec:spline_basics}

In this work, we introduce two adaptive algorithms for simultaneous digital cancellation of multiple even-order \ac{IMD} products, which both rely on spline interpolation to replicate the nonlinear function present in $y_{\text{BB}}^{\text{IMD}}[n]$. As a basis, we summarize the most important properties and definitions of the well-known spline interpolation method, which are then utilized in the derivations in \secref{sec:spline_imdx_canc}.

\subsection{B-Splines}

In many applications, like numerical simulations or computer graphics, it is desired to approximate general nonlinear functions by simpler expressions in order to enable efficient evaluation and easier analysis. Moreover, discrete data series frequently have to be interpolated to obtain intermediate values or to enable analytic manipulation. A straight-forward approximation method for both types of applications is the use of a single polynomial over the whole domain of the target function. While Weierstrass' theorem states that this is generally possible to any desired accuracy, without further precautions oscillations occur (Runge's phenomenon) \cite{Epperson1987,Boor2001}. This effect can be limited by solely using polynomials of low degree. The natural consequence is to employ piecewise polynomial functions, where the approximation accuracy is defined by the number of sections used.

The boundaries of the sections are defined by ${M_{\text{sp}} = N_{\text{sp}} \! + \! Q_{\text{sp}}}$ knots ${[x_0, x_1, \ldots, x_{M_{\text{sp}}-1}]}$, which are sorted in a monotonically increasing order. $N_{\text{sp}}$ is the number of points to interpolate and $Q_{\text{sp}}$ is the order of the spline curve $S(x)$. The curve is composited of polynomial sections of degree ${Q_{\text{sp}} \! - \! 1}$. Depending on the continuity across the knots, the interpolation properties and the support of the base functions, different classes of spline curves are distinguished. We first focus on B-splines \cite{Boor2001,Schumaker2007}, which provide $C^{Q_{\text{sp}}-2}$ smoothness and minimal support, but, in general\footnote{In case of $Q_{\text{sp}} = 1$ (step function) and $Q_{\text{sp}} = 2$ (linear interpolation), the B-spline curve exactly passes through its control points.}, the curve does not exactly interpolate (\ie pass through) its control points $q_i$ \cite{Catmull1974}. $S(x)$ is a linear combination of the base functions $B_{Q_{\text{sp}}, i}(x)$ weighted by $N_{\text{sp}}$ control points $q_i$:
\begin{equation}
	S(x) = \sum_{i = 0}^{N_{\text{sp}}-1} q_i \, B_{Q_{\text{sp}}, i}(x), \qquad x_{Q_{\text{sp}}-1} \leq x < x_{N_{\text{sp}}}.
\end{equation}
The domain of the open spline $S(x)$ is limited compared to the knot vector, because the first ${Q_{\text{sp}} \! - \! 1}$ and the last ${Q_{\text{sp}} \! - \! 1}$ intervals do not have full support. In order to limit the computational effort, we only cover uniform splines, where the lengths of all segments are identical, \ie ${\Delta x = x_{m+1} \! - \! x_m \, \forall m}$. In this case, the control points $q_i$ are located at
\begin{equation}
	\bar{x}_i = x_i + \frac{Q_{\text{sp}}}{2}\Delta x, \quad i = 0,\ldots,N_{\text{sp}}-1.
\end{equation}

The $B_{Q_{\text{sp}}, i}(x)$ can be obtained by the Cox-de Boor recursion \cite{Boor2001}. For illustrative purposes, we provide the explicit forms for ${Q_{\text{sp}} = \{1,2,3\}}$:
{\allowdisplaybreaks
\begin{align}
	B_{1, i}(x) &=
	\begin{cases}
		1 & x_i \leq x < x_{i+1}\\
		0 & \text{else}
	\end{cases}\\
	B_{2, i}(x) &= 
	\begin{cases}
		\frac{1}{\Delta x} \left(x - x_i\right) & x_i \leq x < x_{i+1}\\
		\frac{1}{\Delta x} \left(x_{i+2} - x\right) & x_{i+1} \leq x < x_{i+2}\\
		0 & \text{else}
	\end{cases}\\
	B_{3, i}(x) &= 
	\begin{cases}
		\frac{1}{2 (\Delta x)^2} \left(x - x_i\right)^2 & x_i \leq x < x_{i+1}\\
		\parbox[t]{.45\columnwidth}{$\frac{1}{2 (\Delta x)^2} \big(\!\left(x - x_i\right) \left(x_{i+2} - x\right)$\\ \hspace*{1.4em} $ + \left(x_{i+3} - x\right)\left(x - x_{i+1}\right)\!\big)$}  & x_{i+1} \leq x < x_{i+2}\\
		\frac{1}{2 (\Delta x)^2} \left(x_{i+3} - x\right)^2 & x_{i+2} \leq x < x_{i+3}\\
		0 & \text{else}
	\end{cases}.
\end{align}}%
Since the $B_{Q_{\text{sp}}, i}(x)$ are non-zero only in the interval $[x_i, x_{i + Q_{\text{sp}}})$, a single control point $q_i$ impacts a limited section of the curve. This important property enables an independent adjustment of the $q_i$ by an iterative algorithm. Additionally, the base functions are normalized to form a partition of unity, \ie
\begin{equation}
	\sum_{i = 0}^{Q_{\text{sp}}-1} B_{Q_{\text{sp}}, m-i}(x) = 1, \qquad x_{m} \leq x < x_{m+1}.
	\label{equ:spline_partition_of_unity}
\end{equation}

In order to simplify the evaluation of a uniform spline function, we replace the global input $x$ by the normalized input ${\nu \in [0,1)}$ relative to the lower limit of the interval $\iota$:
\begin{align}
	x &= \nu \Delta x + \iota \Delta x + x_0\\
	\nu &= \frac{x}{\Delta x} - \floor{\frac{x}{\Delta x}}, \quad \iota = \floor{\frac{x - x_0}{\Delta x}}.
\end{align}
This allows to rewrite $S(x)$ into a matrix-vector product, as exemplarily shown for ${Q_{\text{sp}} = 3}$:
\begin{align}
	S(\nu, \iota) &= \sum_{i = 0}^{2} q_{\iota-i} \, B_{3, \iota-i}(\nu \Delta x + \iota \Delta x + x_0) \nonumber\\
	&= \frac{q_{\iota-2}}{2 \left(\Delta x\right)^2} \left(\Delta x\right)^2 (1 - \nu)^2 \nonumber\\
	&\phantom{{}={}} + \frac{q_{\iota-1}}{2 \left(\Delta x\right)^2} \left(\Delta x\right)^2 \left((\nu + 1)(1 - \nu) + (2 - \nu)\nu \right) \nonumber\\
	&\phantom{{}={}} + \frac{q_{\iota}}{2 \left(\Delta x\right)^2} \left(\Delta x\right)^2 \nu^2 \nonumber\\
	&=
	\begin{bmatrix}
		\nu^2 & \nu & 1
	\end{bmatrix}
	\begin{bmatrix}
		\frac{1}{2} & -1 & \frac{1}{2}\\
		-1 & 1 & 0\\
		\frac{1}{2} & \frac{1}{2} & 0
	\end{bmatrix}
	\begin{bmatrix}
		q_{\iota-2}\\
		q_{\iota-1}\\
		q_{\iota}
	\end{bmatrix} \nonumber\\
	&= \vect{\nu}_3\tran \, \matx{B}_3^{\text{sp}} \, \vect{q}_{3, \iota}.
	\label{equ:spline_outp_matrix_notation}
\end{align}
For arbitrary spline order, the input vector $\vect{\nu}_{Q_{\text{sp}}}$ and the control point vector $\vect{q}_{Q_{\text{sp}}, \iota}$ are
\begin{align}
	\vect{\nu}_{Q_{\text{sp}}} &= 
	\begin{bmatrix}
		\nu^{Q_{\text{sp}}-1}, \nu^{Q_{\text{sp}}-2}, \ldots, 1
	\end{bmatrix}\tran \\
	\vect{q}_{Q_{\text{sp}}, \iota} &= 
	\begin{bmatrix}
		q_{\iota - Q_{\text{sp}} + 1}, q_{\iota - Q_{\text{sp}} + 2}, \ldots, q_{\iota}
	\end{bmatrix}\tran.
\end{align}
While the formulation \eqref{equ:spline_outp_matrix_notation} is useful for derivations, in an implementation it is advantageous to calculate \eqref{equ:spline_outp_matrix_notation} using Horner's method, which reduces the required number of multiplications. The B-spline basis matrices $\matx{B}_{Q_{\text{sp}}}^{\text{B}}$ for $Q_{\text{sp}}$ up to 4, \ie cubic interpolation, are
\begin{align}
	&\matx{B}_1^{\text{B}} \coloneqq 1, &&\matx{B}_2^{\text{B}} \coloneqq 
	\begin{bmatrix}
		-1 & 1\\
		1 & 0
	\end{bmatrix}, \label{equ:b_spline_basis_mat_1_2}\\
	&\matx{B}_3^{\text{B}} \coloneqq
	\begin{bmatrix}
		\frac{1}{2} & -1 & \frac{1}{2}\\
		-1 & 1 & 0\\
		\frac{1}{2} & \frac{1}{2} & 0
	\end{bmatrix},
	&&\matx{B}_4^{\text{B}} \coloneqq 
	\begin{bmatrix}
		-\frac{1}{6} & \frac{1}{2} & -\frac{1}{2} & \frac{1}{6}\\
		\frac{1}{2} & -1 & \frac{1}{2} & 0\\
		-\frac{1}{2} & 0 & \frac{1}{2} & 0\\
		\frac{1}{6} & \frac{2}{3} & \frac{1}{6} & 0
	\end{bmatrix}. \label{equ:b_spline_basis_mat_3_4}
\end{align}
The normalization of the basis functions, shown in \eqref{equ:spline_partition_of_unity}, is directly resembled in the matrices $\matx{B}_{Q_{\text{sp}}}$ since we have
\begin{equation}
	\vect{\nu}_{Q_{\text{sp}}}\tran \, \matx{B}_{Q_{\text{sp}}}^{\text{sp}} \, \matx{1}_{Q_{\text{sp}} \times 1} = \vect{\nu}_{Q_{\text{sp}}}\tran \vect{e}_{Q_{\text{sp}}} = 1,
	\label{equ:cr_spline_basis_mat}
\end{equation}
where ${\vect{q}_{Q_{\text{sp}}, \iota} = \matx{1}_{Q_{\text{sp}} \times 1}}$ ensures equal weighting of the base functions, $\matx{1}_{K \times 1}$ is the all-ones vector and $\vect{e}_{k}$ is the $k$-th unit vector.

\subsection{Catmull-Rom Splines}

As a trade-off between approximation accuracy and the risk of overfitting, cubic splines are preferred in many applications. Among all cubic splines, B-splines offer the highest degree of smoothness, but they do not exactly interpolate its control points. Conversely, this means that obtaining the control points of a well approximating spline curve based on a given set of function points necessitates to solve a system of equations. In adaptive learning, however, a simple way to initialize the control points to a given function is beneficial. Catmull-Rom (CR) splines feature the desired interpolation property at the cost of lower smoothness, which is only $C^1$ compared to $C^2$ for cubic B-splines \cite{Catmull1974}. The basis matrix for CR-splines is given by
\begin{equation}
	\matx{B}^{\text{CR}} \coloneqq
	\begin{bmatrix}
		-\frac{1}{2} & \frac{3}{2} & -\frac{3}{2} & \frac{1}{2}\\
		1 & -\frac{5}{2} & 2 & -\frac{1}{2}\\
		-\frac{1}{2} & 0 & \frac{1}{2} & 0\\
		0 & 1 & 0 & 0
	\end{bmatrix}.
\end{equation}
Note that CR-splines are limited to order 4.

\section{Spline-Based Cancellation of Even-Order IMD Self-Interference}
\label{sec:spline_imdx_canc}

An adaptive algorithm used for simultaneous digital cancellation of multiple even-order \ac{IMD} products has to estimate the leakage path $\tilde{h}_{\text{BB}}^{\text{TxL}}[n]$ and the coefficients of the receiver nonlinearity $\gamma_k$. We aim to solve this problem by means of a linear adaptive filter followed by a nonlinear adaptive function based on splines. Hence, in the following we present two novel complex Wiener \ac{SAF} algorithms, which are tailored to the \ac{IMD} effect. Besides the basic algorithm, we cover implementation-related aspects, such as pipelining, and use a constraint optimization technique to avoid internal clipping, an inherent issue of \acp{SAF}.

\subsection{Wiener-SAF with Complex Input, Internal Fixed Nonlinearity and Fixed Output Filter}

\subsubsection{Basic Algorithm}

\begin{figure*}
	\centering
 	\includegraphics[width=0.85\textwidth]{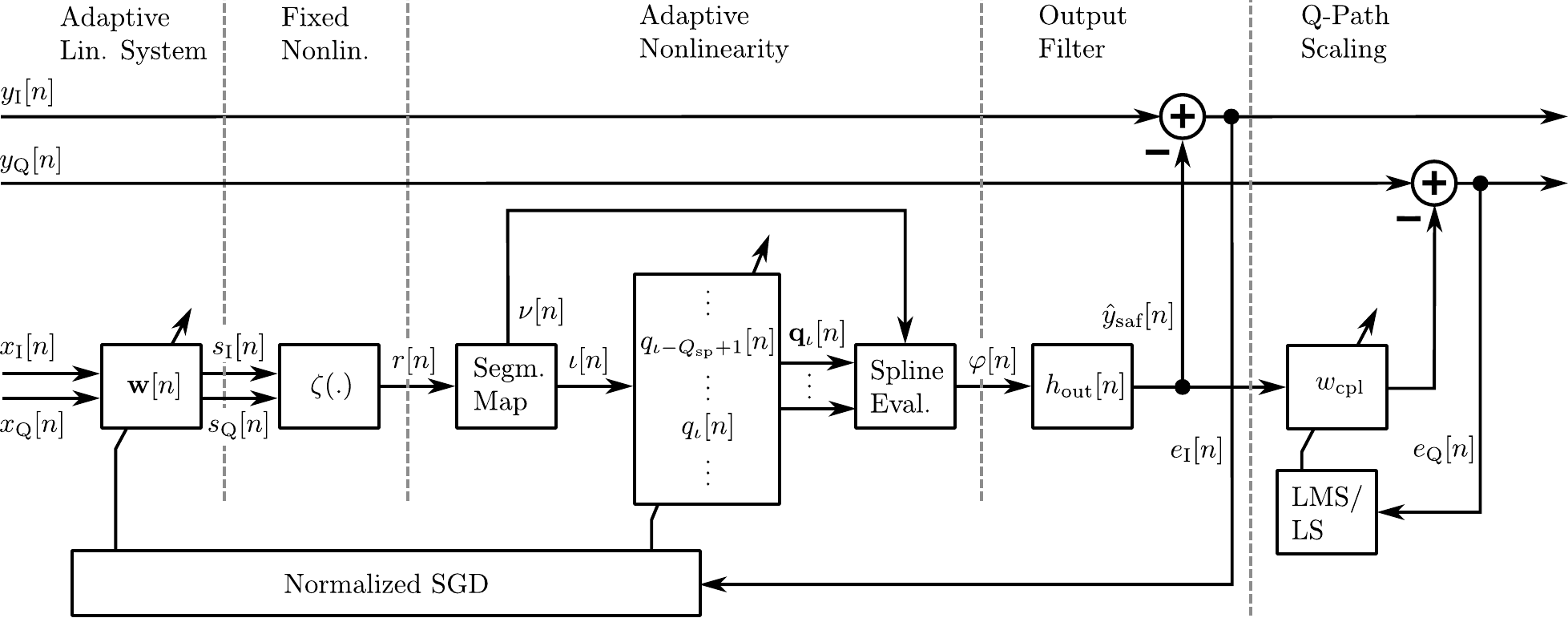}
	\caption{Block diagram of adaptive nonlinear DSIM structure with real output and subsequent single-tap scaling for Q-path.}
	\label{fig:WSAF_ScaledQ}
\end{figure*}%
We propose a Wiener model as shown in the block diagram in \figref{fig:WSAF_ScaledQ}, which employs a fixed nonlinear function ${\zeta: \mathbb{C} \rightarrow \mathbb{R}}$ that transforms the complex filter output $s[n]$ into a real quantity $r[n]$. A suitable function for the \ac{IMD} problem is ${\zeta(s) = |s|^2}$. Alternatively, ${\zeta(s) = |s|}$ could be used, which features an advantageous amplitude distribution. The signal $r[n]$ is the input of the spline function, which is capable of approximating a wide range of nonlinearities. Obviously, this includes the \ac{IMD} products of the form $|.|^{2k}$ and any linear combination of these terms. The latter case matches the interference $y_\text{BB}^{\text{IMD}}[n]$. In the following, we refer to the whole structure as complex-input Wiener SAF (CI-WSAF). Its output ${\hat{y}_{\text{saf}}[n] \in \mathbb{R}}$ is given by
\begin{align}
	\hat{y}_{\text{saf}}[n]	&= \underbrace{\left(\vect{\nu}_{Q_{\text{sp}}}[n]\tran \, \matx{B}_{Q_{\text{sp}}}^{\text{sp}} \, \vect{q}_{Q_{\text{sp}}, \iota[n]}[n \! - \! 1]\right)}_{\varphi[n]}{} * h_{\text{out}}[n] \nonumber\\
	&= \sum_{k = 0}^{Q_{\text{out}} - 1} \varphi[n \! - \! k] \, h_{\text{out}}[k]
\end{align}
using the definitions
\begin{align}
	r[n] &= \zeta(s[n]), \quad s[n] = \vect{w}[n \! - \! 1]\tran \, \vect{x}[n] \label{equ:wsaf_fir_outp_nonlin}\\
	\nu[n] &= \frac{r[n]}{\Delta r} - \floor{\frac{r[n]}{\Delta r}}, \quad \iota[n] = \floor{\frac{r[n] - r_0}{\Delta r}}.
\end{align}
$\hat{y}_{\text{saf}}[n]$ is an approximation of the (noisy) desired sequence $y[n]$, which is $y_{\text{BB,I}}^{\text{Tot}}[n]$ in case of \ac{IMD} cancellation. The corresponding input $x[n]$ of the algorithm is the known \ac{Tx} \ac{BB} stream $x_{\text{BB}}[n]$. The real impulse response $h_{\text{out}}[n]$ allows to apply a known filter to the output of the spline function. Depending on the setup, this filter could be a \ac{CSF} or a pure delay. The latter is important to model pipelining stages in the output computation, which help to achieve the desired operating frequency. The length of the adaptive filter $\vect{w}[n]$ is $Q_{\text{lin}}$, the length of the output filter $h_{\text{out}}[n]$ is $Q_{\text{out}}$ and the spline order is $Q_{\text{sp}}$. We use uniform knots with a spacing of $\Delta r$ and a lower limit of $r_0$. When calculating $\iota[n]$, it is important to realize that its value must be in the range $[Q_{\text{sp}} - 1, N_{\text{sp}} - 1]$, where $N_{\text{sp}}$ is the number of control points. For any values outside of this range, the correct spline output cannot be calculated due to an index underrun or overflow in $\vect{q}_{Q_{\text{sp}}, \iota}$. The vectors $\vect{\nu}_{Q_{\text{sp}}}[n]$, $\vect{q}_{Q_{\text{sp}}, \iota[m]}[n]$, $\vect{w}[n]$ and $\vect{x}[n]$ are given by
\begin{align}
	\vect{\nu}_{Q_{\text{sp}}}[n] &= 
	\begin{bmatrix}
		\left(\nu[n]\right)^{Q_{\text{sp}}-1}, \left(\nu[n]\right)^{Q_{\text{sp}}-2}, \ldots, 1
	\end{bmatrix}\tran \\
	\vect{q}_{Q_{\text{sp}}, \iota[m]}[n] &= 
	\begin{bmatrix}
		q_{\iota[m] - Q_{\text{sp}} + 1}[n], q_{\iota[m] - Q_{\text{sp}} + 2}[n], \ldots, q_{\iota[m]}[n]
	\end{bmatrix}\tran\\
	\vect{w}[n] &=
	\begin{bmatrix}
		w_0[n], w_1[n], \ldots, w_{Q_{\text{lin}}-1}[n]
	\end{bmatrix}\tran\\
	\vect{x}[n] &=
	\begin{bmatrix}
		x[n], x[n \! - \! 1], \ldots, x[n \! - \! Q_{\text{lin}} \! + \! 1]
	\end{bmatrix}\tran.
\end{align}
Note the difference in the time indices between the segment index $\iota[m]$ and the control points $q_{\iota}[n]$. The $q_{\iota}[n]$ and thus also $\hat{y}[n]$ are real-valued. For the following derivations, we also define the vector ${\vect{q}[n] = [q_0[n], q_1[n], \ldots, q_{N_{\text{sp}}-1}[n]]\tran}$ comprising all control points, and use the indexing scheme ${q_{\iota}[n] = [\vect{q}[n]]_{\iota}}$.

We propose to estimate the parameters $\vect{w}$ and $\vect{q}$ by means of \iac{SGD} method. The corresponding cost function is the instantaneous squared estimation error, augmented by an additional penalty using the $p$-norm of the weights:
\begin{align}
	J[n] &= \abs{e[n]}^2 + \epsilon \, \underbrace{\left(\norm{\vect{w}[n \! - \! 1]}_p^p - \rho_{w}\right)^2}_{c_{\vect{w}}[n]}\\
	e[n] &= y[n] - \hat{y}_{\text{saf}}[n].
	\label{equ:wsaf_err}
\end{align}
$\rho_{w}$ is the target value for $\norm{\vect{w}[n]}_p^p$ and $\epsilon$ is a weighting factor. The norm constrained \ac{SGD} concept is known from sparse estimation, where, for instance, an approximation of the $\ell_0$ norm is employed \cite{Gu2009}. In the \ac{SAF} application, the goal is to avoid any gain ambiguity between the weights and the control points on the one hand and to ensure a limited number range at the input of the spline function on the other hand. This avoids saturation of the signal $r[n]$, which could lead to instability of the algorithm. From \ac{FIR} filter implementations in fixed-point arithmetic it is known that the $\ell_1$ norm of the weights poses a conservative bound on the output range for a given input number format \cite{Gruenigen2014}. A computationally less expensive approximation would be the $\ell_2$ norm, where it holds that ${\norm{\vect{w}}_2 \leq \norm{\vect{w}}_1}$. Thus, the $\ell_2$ norm is less restrictive and does not guarantee that overflows are eliminated completely. In order to minimize the cost function, we require the gradients of $J[n]$ with respect to the parameters.

Since the differentiation is not directly possible due to the time-delays caused by the convolution with $h_{\text{out}}[n]$, we introduce approximations, where the targeted parameter is assumed to be time-independent. Consequently, the approximated gradient with respect to $\vect{q}$ is
\begin{align}
	\tilde{\vect{g}}_{\vect{q}}[n]\tran &= \frac{\partial J[n]}{\partial e[n]} \frac{\partial e[n]\vert_{\vect{q}[.] = \vect{q}}}{\partial \vect{q}} \nonumber\\
	&= 2 \, e[n] \, \tilde{\vect{g}}_{e,\vect{q}}[n]\tran.
\end{align}
Throughout this work, any derivative of the form $\partial f(\vect{x})/\partial \vect{x}$ shall be a row vector when $\vect{x}$ is a column vector \cite{Dhrymes2000}. This allows for a straight-forward application of the chain rule. When evaluating the partial derivative $\tilde{\vect{g}}_{e,\vect{q}}[n]$ we yield
\begin{align}
	\tilde{\vect{g}}_{e,\vect{q}}[n] &= -\sum_{k = 0}^{Q_{\text{out}} - 1} h_{\text{out}}[k]
	\begin{bmatrix}
		\matx{0}_{(\iota[n - k] - Q_{\text{sp}} + 1) \times 1}\\
		(\matx{B}_{Q_{\text{sp}}}^{\text{sp}})\tran \, \vect{\nu}_{Q_{\text{sp}}}[n \! - \! k]\\
		\matx{0}_{(N_{\text{sp}} - \iota[n - k] - 1) \times 1}
	\end{bmatrix},
	\label{equ:wsaf_grad_ctrl_pts}
\end{align}
where the all-zero vectors $\matx{0}_{K \times 1}$ are used to appropriately place the derivative of the spline output depending on the interval index at the time step ${n \! - \! k}$. Since the evaluation of \eqref{equ:wsaf_grad_ctrl_pts} might be computationally expensive depending on $Q_{\text{out}}$, a possible approximation is to neglect small $h_{\text{out}}[k]$ or to replace the filter with its gain and group delay. The approximated gradient with respect to $\vect{w}$ is obtained in a similar manner
\begin{align}
	\tilde{\vect{g}}_{\vect{w}}[n]\tran &= \frac{\partial J[n]}{\partial e[n]} \frac{\partial e[n]\vert_{\vect{w}[.] = \vect{w}}}{\partial \vect{w}^*} + \frac{\partial J[n]}{\partial c_{\vect{w}}[n]} \frac{c_{\vect{w}}[n]}{\partial \vect{w}[n \! - \! 1]^*} \nonumber\\
	&= 2 \, e[n] \, \tilde{\vect{g}}_{e,\vect{w}}[n]\tran + \epsilon \, \tilde{\vect{g}}_{c,\vect{w}}[n]\tran
\end{align}
using the $\mathbb{CR}$ (or Wirtinger) calculus \cite{Remmert1991}, which conveniently guarantees the correct direction of the gradient for both, the real and the imaginary part of $\vect{w}$. The substitution of $\vect{w}[.]$ shall affect only the spline input $\nu[.]$, but not the spline segment indices $\iota[.]$. We assume that $\zeta(s, s^*)$ and $c_{\vect{w}}(\vect{w}, \vect{w}^*)$ fulfill Brandwood's analyticity condition \cite{Brandwood1983}, thereby simplifying the $\mathbb{CR}$ derivatives. The term $\tilde{\vect{g}}_{e,\vect{w}}[n]$ evaluates to
\begin{align}
	&\tilde{\vect{g}}_{e,\vect{w}}[n]\nonumber\\
	&= \sum_{k = 0}^{Q_{\text{out}} - 1} \frac{\partial e[n]}{\partial \vect{\nu}_{Q_{\text{sp}}}[n \! - \! k]} \frac{\partial \vect{\nu}_{Q_{\text{sp}}}[n \! - \! k]}{\partial \nu[n \! - \! k]} \nonumber\\
	&\phantom{{} = {}} \hspace{4em}\frac{\partial \nu[n \! - \! k]}{\partial r[n \! - \! k]} \frac{\partial r[n \! - \! k]}{\partial s[n \! - \! k]^*} \frac{\partial s[n \! - \! k]^*\vert_{\vect{w}[.] = \vect{w}}}{\partial \vect{w}\herm} \nonumber\\
	&= -\frac{1}{\Delta r} \sum_{k = 0}^{Q_{\text{out}} - 1} h_{\text{out}}[k] \, \vect{\nu}_{Q_{\text{sp}}}'[n \! - \! k]\tran \matx{B}_{Q_{\text{sp}}}^{\text{sp}} \vect{q}_{Q_{\text{sp}}, \iota[n - k]}[n \! - \! 1 \! - \! k] \nonumber\\
	&\phantom{{} = {}} \hspace{5.5em} \zeta'(s[n \! - \! k])^* \, \vect{x}[n \! - \! k]^*
\end{align}
with ${\zeta'(s) = \partial \zeta(s) / \partial s}$ and
\begin{align}
	&\vect{\nu}_{Q_{\text{sp}}}'[n] = \nonumber\\
	&\hspace{2.5em}\begin{bmatrix}
	\left(Q_{\text{sp}} \! - \! 1\right)(\nu[n])^{Q_{\text{sp}}-2}, \left(Q_{\text{sp}} \! - \! 2\right)(\nu[n])^{Q_{\text{sp}}-3}, \ldots, 0
	\end{bmatrix}\tran.
\end{align}
When using ${\zeta(s) = |s|^2}$, the $\mathbb{CR}$ derivative of the fixed nonlinearity is ${\zeta'(s) = s^*}$. Within one spline segment, the index $\iota[.]$ is constant with respect to $\vect{w}$ due to the floor function. Again, small values of $h_{\text{out}}[k]$ could be neglected to reduce the complexity of evaluating $\tilde{\vect{g}}_{e,\vect{w}}[n]$. The derivative of the norm constraint $\tilde{\vect{g}}_{c,\vect{w}}[n]$ depends on the chosen norm $p$:
\begin{equation}
	\tilde{\vect{g}}_{c,\vect{w}}[n] =
	\begin{dcases}
		\left(\norm{\vect{w}[n \! - \! 1]}_1 - \rho_w\right) & p = 1\\
		\hspace{5em} \left[\frac{w_k[n \! - \! 1]}{\abs{w_k[n \! - \! 1]}}\right]_{k \, = \, 0,\ldots,Q_{\text{lin}}-1} & \\
		2 \left(\norm{\vect{w}[n \! - \! 1]}_2^2 - \rho_w\right) \vect{w}[n \! - \! 1] & p = 2
	\end{dcases}
	\label{equ:wsaf_grad_norm_constr_weights}
\end{equation}
with ${w_k = [\vect{w}]_k}$. In an implementation it might be beneficial to use the equivalence ${w_k/\abs{w_k} = e^{\imag \arg(w_k)}}$. Combining the above results, the update equations for iterative optimization of the parameters are given by
\begin{align}
	\vect{q}[n] & = \vect{q}[n \! - \! 1] - 2\, \tau \, \mu[n] \, e[n] \, \tilde{\vect{g}}_{e,\vect{q}}[n]\\
	\vect{w}[n] &= \vect{w}[n \! - \! 1] - \mu[n] \left(2\, e[n] \, \tilde{\vect{g}}_{e,\vect{w}}[n] + \epsilon \, \tilde{\vect{g}}_{c,\vect{w}}[n]\right)
\end{align}
using the time-dependent non-negative step-size ${\mu[n] \in \mathbb{R}_{\geq 0}}$ and the non-negative coupling factor ${\tau \in \mathbb{R}_{\geq 0}}$.

\vspace{5pt}
\subsubsection{Step-Size Normalization}

The standard \ac{SGD} approach is difficult to tune, since the effective learning rate depends on the dynamics of the involved signals. Thus, commonly a normalized variant is used. One method to derive the normalization is to consider a partial Taylor series expansion of the error signal \cite{Hanna2003, Scarpiniti2014}, which is truncated after the linear term
\begin{align}
	e[&n \! + \! 1] \nonumber\\
	&\approx e[n] + \frac{\partial e[n]\vert_{\vect{q}[.] = \vect{q}}}{\partial \vect{q}} \Delta \vect{q}[n] \nonumber\\
	& \phantom{{} \approx {}} + \frac{\partial e[n]\vert_{\vect{w}[.] = \vect{w}}}{\partial \vect{w}} \Delta \vect{w}[n] + \frac{\partial e[n]\vert_{\vect{w}[.] = \vect{w}}}{\partial \vect{w}^*} \Delta \vect{w}[n]^* \nonumber\\
	&\approx e[n] - 2 \, \tau \, \mu[n] \, e[n] \norm{\tilde{\vect{g}}_{e, \vect{q}}[n]}_2^2 - 4 \, \mu[n] \, e[n] \norm{\tilde{\vect{g}}_{e,\vect{w}}[n]}_2^2.
\end{align}
$\Delta \vect{q}[n]$ and $\Delta \vect{w}[n]$ are the parameter changes from time step ${n \! - \! 1}$ to $n$. Assuming $\epsilon$ to be small, we neglect the weight norm constraint in the normalization. Based on the Taylor expansion, it has been shown that the adaptation performance of \ac{SGD} algorithms is improved if the step-size is chosen to fulfill the following condition:
\begin{align}
	\abs{e[n]} &\geq \abs{e[n \! + \! 1]} \nonumber\\
	\abs{e[n]} &\geq \abs{e[n]} \abs{1 - 2 \, \tau \, \mu[n] \norm{\tilde{\vect{g}}_{e, \vect{q}}[n]}_2^2 - 4 \, \mu[n] \norm{\tilde{\vect{g}}_{e,\vect{w}}[n]}_2^2} \nonumber\\
	1 &\geq \abs{1 - 2 \, \mu[n] \left(\tau \norm{\tilde{\vect{g}}_{e, \vect{q}}[n]}_2^2 + 2 \norm{\tilde{\vect{g}}_{e,\vect{w}}[n]}_2^2\right)}.
\end{align}
Solving the inequality for $\mu[n]$ yields
\begin{equation}
	0 \leq \mu[n] \leq \frac{1}{2 \norm{\tilde{\vect{g}}_{e,\vect{w}}[n]}_2^2 + \tau \norm{\tilde{\vect{g}}_{e, \vect{q}}[n]}_2^2}
\end{equation}
or alternatively
\begin{equation}
	\mu[n] = \frac{\mu}{2 \norm{\tilde{\vect{g}}_{e,\vect{w}}[n]}_2^2 + \tau \norm{\tilde{\vect{g}}_{e, \vect{q}}[n]}_2^2 + \xi}, \quad 0 \leq \mu \leq 1,
\end{equation}
where the regularization value $\xi$ places an upper bound on $\mu[n]$. The maximum adaptation rate is obtained for ${\mu = 0.5}$, higher values up to 1 will slow down the optimization again. While the normalization concludes the derivation of the basic CI-WSAF, we proceed by discussing several useful modifications and extensions.

\vspace{5pt}
\subsubsection{Complex Output}

In the \ac{IMD} problem, a complex-valued interference signal ${y_{\text{BB}}^{\text{IMD}}[n] = y_{\text{BB,I}}^{\text{IMD}}[n] + \imag y_{\text{BB,Q}}^{\text{IMD}}[n]}$ has to be replicated, but the CI-WSAF only provides a real-valued output. When we assume nonlinearity coefficients of the form ${\gamma_k = \gamma_{k, \text{I}} \, (1 + \imag \delta_{\text{Q}})}$ with ${\gamma_{k, \text{I}}, \delta_{\text{Q}} \in \mathbb{R}}$, then the imaginary part of $y_{\text{BB}}^{\text{IMD}}[n]$ is just a scaled copy of the real part. The coupling $\delta_{\text{Q}}$ might be estimated by using a single-tap normalized least mean squares (N-LMS)%
\acuse{LMS}%
\acuse{N-LMS}
or a weighted \ac{LS} algorithm, following a cascaded scheme as illustrated in \figref{fig:WSAF_ScaledQ}. The interference replica used for cancellation is
\begin{equation}
	\hat{y}_{\text{BB}}^{\text{IMD}}[n] = \hat{y}_{\text{saf}}[n] + \imag w_{\text{cpl}}[n] \, \hat{y}_{\text{saf}}[n]
	\label{equ:wsaf_1tap_q_scaling}
\end{equation}
with the weight calculations
\begin{equation}
	w_{\text{cpl}}[n] = w_{\text{cpl}}[n \! - \! 1] + \frac{\mu_{\text{cpl}}}{\left(\hat{y}_{\text{saf}}[n]\right)^2 \! + \! \xi} e_{\text{Q}}[n] \hat{y}_{\text{saf}}[n], \  0 \leq \mu_{\text{cpl}} \leq 1	
\end{equation}
for the 1-tap \ac{N-LMS} using ${e_{\text{Q}}[n] = y_{\text{Q}}[n] - w_{\text{cpl}}[n]\, \hat{y}_{\text{saf}}[n]}$ or
\begin{align}
	w_{\text{cpl}}[n] &= \frac{r_{y\hat{y}}[n]}{r_{\hat{y}\hat{y}}[n]} \label{equ:wsaf_1_tap_weighted_ls}\\
	r_{y\hat{y}}[n] &= r_{y\hat{y}}[n \! - \! 1] + \lambda_{\text{cpl}} \, y_{\text{Q}}[n] \, \hat{y}_{\text{saf}}[n], &r_{y\hat{y}}[-1] = 0\\
	r_{\hat{y}\hat{y}}[n] &= r_{\hat{y}\hat{y}}[n \! - \! 1] + \lambda_{\text{cpl}} \left(\hat{y}_{\text{saf}}[n]\right)^2, &r_{\hat{y}\hat{y}}[-1] = 0
\end{align}
for the exponentially weighted single-tap \ac{LS} algorithm.

\vspace{5pt}
\subsubsection{Transform-Domain Concept}

Furthermore, it is well-known that \ac{SGD} algorithms suffer from slow convergence in case of correlated input signals \cite{Diniz2002}. This property is particularly critical if the reference signal is \iac{LTE}/\ac{NR} sequence with narrow allocation, a common case for uplink. In \cite{Motz2020_C2}, it has been shown that the \ac{TD} concept provides an appropriate mitigation if a precomputed power normalization is available. For the CI\nobreakdash-WSAF, this approach is applied by replacing the input vector $\vect{x}[n]$ with the transformed vector $\vect{v}[n]$:
\begin{equation}
	\vect{v}[n] = \matx{P}^{-\frac{1}{2}} \bm{\mathcal{D}}_{Q_{\text{lin}}} \vect{x}[n], \quad [\matx{P}]_{k,l} = \delta_{k,l} \left[\bm{\mathcal{D}}_{Q_{\text{lin}}} \matx{C}_{\vect{x}\vect{x}} \bm{\mathcal{D}}_{Q_{\text{lin}}}\tran\right]_{k,l}.
	\label{equ:wsaf_td_concept}
\end{equation}
${\bm{\mathcal{D}}_{Q_{\text{lin}}} \in \mathbb{R}^{Q_{\text{lin}} \times Q_{\text{lin}}}}$ is the \ac{DCT} matrix and $\delta_{k,l}$ is the Kronecker delta. $\matx{C}_{\vect{x}\vect{x}}$ is the auto-covariance matrix of the input vector $\vect{x}[n]$, which is used to compute the power normalization $\matx{P}^{-\frac{1}{2}}$. When calculating $\matx{C}_{\vect{x}\vect{x}}$, it is important to take any resampling of $x[n]$ into account. Besides operating on a different input, the CI-WSAF algorithm in the \ac{TD} variant stays unaltered.

\vspace{5pt}
\subsubsection{Weight Norm Limiting}

Last, in the iterative optimization of $\vect{w}[n]$, the norm correction ${\epsilon \, \tilde{\vect{g}}_{c,\vect{w}}[n]}$ has to be applied in every iteration. In order to relax the requirements on a hardware implementation, we propose an alternative heuristic rescaling of the weights
\begin{equation}
	\vect{w}_{\text{lim}}[n] =
	\begin{dcases}
		\vect{w}[n] & \norm{\vect{w}[n \! - \! 1]}_p^p < \rho_w\\
		\frac{\vect{w}[n]}{2} & \text{else}
	\end{dcases},
	\label{equ:wsaf_norm_limiter}
\end{equation}
where the decision can be computed in parallel to the update step and the potential rescaling is a simple shift operation. Obviously, this scaling has to be compensated by the spline control points $\vect{q}$ in order to maintain the output signal level. According to simulations, the iterative adaptation of $\vect{q}$ is capable of this task if the rescaling is applied seldomly (cf.\ \figref{fig:Constr_SGD_Comp}).

\subsection{Extension to Complex Control Points}

\begin{figure*}
	\centering
	\includegraphics[width=0.82\textwidth]{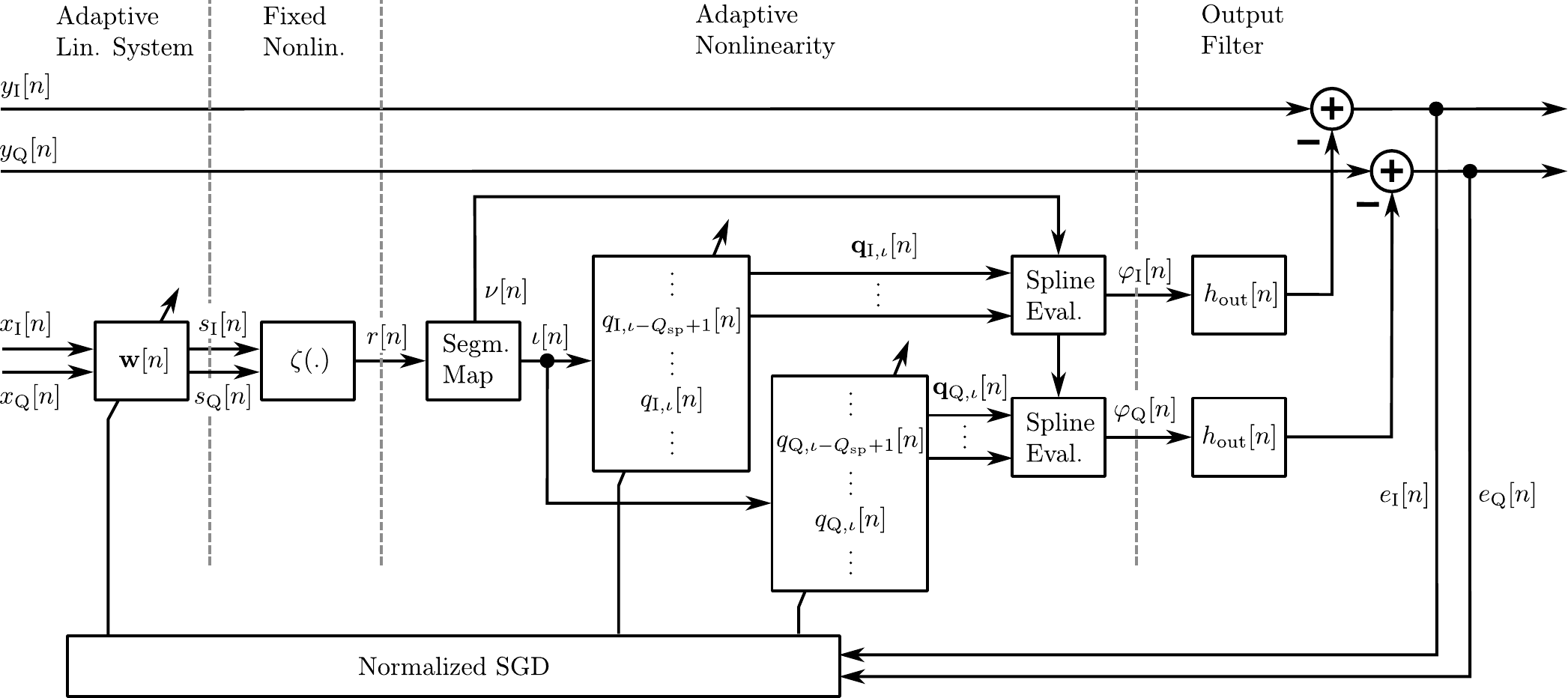}
	\caption{Block diagram of adaptive nonlinear DSIM structure with complex output.}
	\label{fig:WSAF_indepIQ}
\end{figure*}%
The CI-WSAF presented above operates on complex input data and uses complex filter weights, but in its basic form has a real-valued output. Still, a complex interference can be canceled by means of an additional single-tap adaptive filter, under the assumption that the interference in the Q-path is a scaled version of the I-path. When considering the \ac{IMD} \ac{BB} model \eqref{equ:imdx_mdl_aaf_outp}, this constraint is not fulfilled in general. Therefore, we extend the algorithm by including separate spline functions for the I- and the Q-path.

\vspace{5pt}
\subsubsection{Basic Algorithm}

The structure of this algorithm is illustrated in \figref{fig:WSAF_indepIQ}. The output of the so-called complex-input-output Wiener SAF (CIO-WSAF) is given by
\begin{align}
	\hat{y}_{\text{saf}}[n] &= \Big(\vect{\nu}_{Q_{\text{sp}}}[n]\tran \, \matx{B}_{Q_{\text{sp}}}^{\text{sp}} \nonumber\\
	&\hspace{2.5em}\left(\vect{q}_{\text{I}, Q_{\text{sp}}, \iota[n]}[n \! - \! 1] + \imag \vect{q}_{\text{Q}, Q_{\text{sp}}, \iota[n]}[n \! - \! 1]\right)\Big) * h_{\text{out}}[n] \nonumber\\
	&= \underbrace{\left(\vect{\nu}_{Q_{\text{sp}}}[n]\tran \, \matx{B}_{Q_{\text{sp}}}^{\text{sp}} \, \vect{q}_{Q_{\text{sp}}, \iota[n]}[n \! - \! 1]\right)}_{\varphi[n]}{} * h_{\text{out}}[n],
\end{align}
where ${\vect{q}[n] \in \mathbb{C}^{N_{\text{sp}}}}$ and ${\hat{y}_{\text{saf}}[n] \in \mathbb{C}}$. Despite this change, all signal definitions and the \ac{SGD} cost function, \eqref{equ:wsaf_fir_outp_nonlin}--\eqref{equ:wsaf_err}, remain valid. In the \ac{IMD} problem, the output of the CIO\nobreakdash-WSAF can be directly used for interference cancellation, \ie ${\hat{y}_{\text{BB}}^{\text{IMD}}[n] = \hat{y}_{\text{saf}}[n]}$. The approximate gradient of the cost function with respect to the control points now also requires the Wirtinger calculus. Applying the corresponding chain rule leads to:
\begin{align}
	\tilde{\vect{g}}_{\vect{q}}[n] &= \frac{\partial J[n]}{\partial e[n]} \frac{\partial e[n]\vert_{\vect{q}[.] = \vect{q}}}{\partial \vect{q}\herm} + \frac{\partial J[n]}{\partial e[n]^*} \frac{\partial e[n]^*\vert_{\vect{q}^*[.] = \vect{q}^*}}{\partial \vect{q}\herm} \nonumber\\
	&= - e[n] \sum_{k = 0}^{Q_{\text{out}} - 1} h_{\text{out}}[k]
	\begin{bmatrix}
		\matx{0}_{(\iota[n - k] - Q_{\text{sp}} + 1) \times 1}\\
		(\matx{B}_{Q_{\text{sp}}}^{\text{sp}})\tran \, \vect{\nu}_{Q_{\text{sp}}}[n \! - \! k]\\
		\matx{0}_{(N_{\text{sp}} - \iota[n - k] - 1) \times 1}
	\end{bmatrix}\nonumber\\
	&= e[n] \, \tilde{\vect{g}}_{e,\vect{q}}[n].
\end{align}
Due to the real spline input, this result can be interpreted as an independent optimization of the nonlinearities in the I- and the Q-path. The same structure of the chain rule also occurs in case of the approximate gradient with respect to the filter weights:
\begin{align}
		\tilde{\vect{g}}_{\vect{w}}[n]\tran &= \frac{\partial J[n]}{\partial e[n]} \frac{\partial e[n]\vert_{\vect{w}[.] = \vect{w}}}{\partial \vect{w}^*} + \frac{\partial J[n]}{\partial e[n]^*} \frac{\partial e[n]^*\vert_{\vect{w}[.] = \vect{w}}}{\partial \vect{w}^*}\nonumber\\
		&\phantom{{} = {}} + \frac{\partial J[n]}{\partial c_{\vect{w}}[n]} \frac{c_{\vect{w}}[n]}{\partial \vect{w}[n \! - \! 1]^*} \nonumber\\
		&= \tilde{\vect{g}}_{J,e,\vect{w}}[n]\tran + \epsilon \, \tilde{\vect{g}}_{c,\vect{w}}[n]\tran.
\end{align}
The notation $\tilde{\vect{g}}_{J,e,\vect{w}}[n]$ indicates that no separation between the error $e[n]$ and the partial derivative of $e[n]$ is possible anymore, which is a major difference to the real case. The gradient $\tilde{\vect{g}}_{J,e,\vect{w}}[n]$ is given by
\begin{align}
	&\tilde{\vect{g}}_{J,e,\vect{w}}[n]\nonumber\\
	&= \frac{\partial J[n]}{\partial e[n]} \sum_{k = 0}^{Q_{\text{out}} - 1} \frac{\partial e[n]}{\partial \vect{\nu}_{Q_{\text{sp}}}[n \! - \! k]} \frac{\partial \vect{\nu}_{Q_{\text{sp}}}[n \! - \! k]}{\partial \nu[n \! - \! k]} \nonumber\\
	&\hspace{7em}\frac{\partial \nu[n \! - \! k]}{\partial r[n \! - \! k]} \frac{\partial r[n \! - \! k]}{\partial s[n \! - \! k]^*} \frac{\partial s[n \! - \! k]^*\vert_{\vect{w}[.] = \vect{w}}}{\partial \vect{w}\herm} \nonumber\\
	&\phantom{{} = {}} + \frac{\partial J[n]}{\partial e[n]^*} \sum_{k = 0}^{Q_{\text{out}} - 1} \frac{\partial e[n]^*}{\partial \vect{\nu}_{Q_{\text{sp}}}[n \! - \! k]} \frac{\partial \vect{\nu}_{Q_{\text{sp}}}[n \! - \! k]}{\partial \nu[n \! - \! k]} \nonumber\\
	&\hspace{8.5em}\frac{\partial \nu[n \! - \! k]}{\partial r[n \! - \! k]} \frac{\partial r[n \! - \! k]}{\partial s[n \! - \! k]^*} \frac{\partial s[n \! - \! k]^*\vert_{\vect{w}[.] = \vect{w}}}{\partial \vect{w}\herm} \nonumber\\
	&= -\frac{2}{\Delta r} \sum_{k = 0}^{Q_{\text{out}} - 1} h_{\text{out}}[k] \, \vect{\nu}_{Q_{\text{sp}}}'[n \! - \! k]\tran \matx{B}_{Q_{\text{sp}}}^{\text{sp}} \nonumber\\
	&\hspace{3.3em} \Re\!\left\{e[n]^* \, \vect{q}_{Q_{\text{sp}}, \iota[n - k]}[n \! - \! 1 \! - \! k]\right\} \, \zeta'(s[n \! - \! k])^* \, \vect{x}[n \! - \! k]^*.
\end{align}
This form is obtained by factoring out all common terms and using the identity ${a + a^* = 2 \, \Re\{a\}}$. The gradient of the norm constraint \eqref{equ:wsaf_grad_norm_constr_weights}, as well as the limiter \eqref{equ:wsaf_norm_limiter}, remain unaffected by the complex control points and can be directly reused. Combining all results, the update equations of the CIO-WSAF are
\begin{align}
	\vect{q}[n] & = \vect{q}[n \! - \! 1] - \tau \, \mu[n] \, e[n] \, \tilde{\vect{g}}_{e,\vect{q}}[n]\\
	\vect{w}[n] &= \vect{w}[n \! - \! 1] - \mu[n] \left(\tilde{\vect{g}}_{J,e,\vect{w}}[n] + \epsilon \, \tilde{\vect{g}}_{c,\vect{w}}[n]\right).
\end{align}

\vspace{5pt}
\subsubsection{Step-Size Normalization}

Similar to the real case, the adaptation performance of the algorithm is improved by calculating a partial linear Taylor approximation of the error signal and choosing the step-size such that the error decreases with each iteration. In order to simplify the corresponding inequality for $\mu[n]$, we replace the convolution with $h_{\text{out}}[n]$ by the group delay $k_{\text{g}}$ and the passband gain $h_{\text{g}}$ of the output filter.
\begin{align}
	&e[n \! + \! 1] \nonumber\\
	&\approx e[n] + \frac{\partial e[n]\vert_{\vect{q}[.] = \vect{q}}}{\partial \vect{q}} \Delta \vect{q}[n] \nonumber\\
	& \phantom{{} \approx {}} + \frac{\partial e[n]\vert_{\vect{w}[.] = \vect{w}; \, h_{\text{g}}}}{\partial \vect{w}} \Delta \vect{w}[n] + \frac{\partial e[n]\vert_{\vect{w}^*[.] = \vect{w}^*; \, h_{\text{g}}}}{\partial \vect{w}^*} \Delta \vect{w}[n]^* \nonumber\\
	&\approx e[n] - \tau \, \mu[n] \, e[n] \norm{\tilde{\vect{g}}_{e, \vect{q}}[n]}_2^2 \nonumber\\
	&\phantom{{} \approx {}} - \mu[n] \, \frac{4}{(\Delta r)^2} \, h_{\text{g}}^2 \, \vect{\nu}_{Q_{\text{sp}}}'[n \! - \! k_{\text{g}}]\tran \matx{B}_{Q_{\text{sp}}}^{\text{sp}} \vect{q}_{Q_{\text{sp}}, \iota[n - k_{\text{g}}]}[n \! - \! 1 \! - \! k_{\text{g}}]\nonumber\\
	&\phantom{{} \approx {} - {}} \vect{\nu}_{Q_{\text{sp}}}'[n \! - \! k_{\text{g}}]\tran \matx{B}_{Q_{\text{sp}}}^{\text{sp}} \Re\!\left\{e[n]^* \, \vect{q}_{Q_{\text{sp}}, \iota[n - k_{\text{g}}]}[n \! - \! 1 \! - \! k_{\text{g}}]\right\} \nonumber\\
	&\phantom{{} \approx {} - {}} \abs{\zeta'(s[n \! - \! k_{\text{g}}])}^2 \norm{\vect{x}[n \! - \! k_{\text{g}}]}_2^2
\end{align}
In the classical derivation of the step-size bound, it is assumed that ${\abs{e[n\!+\!1]} \leq \abs{e[n]}}$. Since no closed-form solution of this inequality for $\mu[n]$ independent of $e[n]$ is possible, we instead place constraints in the real and imaginary part of $e[n]$ separately:
\begin{align}
	\abs{e_{\text{I}}[n]} &\geq \abs{e_{\text{I}}[n \! + \! 1]} \nonumber\\
	&\geq \bigg\vert e_{\text{I}}[n] - \tau \, \mu[n] \, e_{\text{I}}[n] \norm{\tilde{\vect{g}}_{e, \vect{q}}[n]}_2^2 \nonumber\\
	&\phantom{{} \geq {}} - \mu[n] \, e_{\text{I}}[n] \, \frac{4}{(\Delta r)^2} \, h_{\text{g}}^2 \, \abs{\zeta'(s[n \! - \! k_{\text{g}}])}^2 \norm{\vect{x}[n \! - \! k_{\text{g}}]}_2^2 \nonumber\\
	&\phantom{{} \geq {} - {}} \left(\vect{\nu}_{Q_{\text{sp}}}'[n \! - \! k_{\text{g}}]\tran \matx{B}_{Q_{\text{sp}}}^{\text{sp}} \vect{q}_{\text{I}, Q_{\text{sp}}, \iota[n - k_{\text{g}}]}[n \! - \! 1 \! - \! k_{\text{g}}]\right)^2 \bigg\vert \nonumber\\
	&\geq \abs{e_{\text{I}}[n]} \abs{1 - \mu[n] \, b_{\mu,\text{I}}[n]}
	\label{equ:cio_wsaf_normalization_ei}
\end{align}
\begin{align}
	\abs{e_{\text{Q}}[n]} &\geq \abs{e_{\text{Q}}[n \! + \! 1]} \nonumber\\
	&\geq \bigg\vert e_{\text{Q}}[n] - \tau \, \mu[n] \, e_{\text{Q}}[n] \norm{\tilde{\vect{g}}_{e, \vect{q}}[n]}_2^2 \nonumber\\
	&\phantom{{} \geq {}} - \mu[n] \, e_{\text{Q}}[n] \, \frac{4}{(\Delta r)^2} \, h_{\text{g}}^2 \, \abs{\zeta'(s[n \! - \! k_{\text{g}}])}^2 \norm{\vect{x}[n \! - \! k_{\text{g}}]}_2^2 \nonumber\\
	&\phantom{{} \geq {} - {}} \left(\vect{\nu}_{Q_{\text{sp}}}'[n \! - \! k_{\text{g}}]\tran \matx{B}_{Q_{\text{sp}}}^{\text{sp}} \vect{q}_{\text{Q}, Q_{\text{sp}}, \iota[n - k_{\text{g}}]}[n \! - \! 1 \! - \! k_{\text{g}}]\right)^2 \bigg\vert \nonumber\\
	&\geq \abs{e_{\text{Q}}[n]} \abs{1 - \mu[n] \, b_{\mu,\text{Q}}[n]}.
	\label{equ:cio_wsaf_normalization_eq}
\end{align}
Here, we used the definitions
\begin{align}
	&b_{\mu,\text{I}}[n] \nonumber\\
	&\hspace{0.5em} = \tau \norm{\tilde{\vect{g}}_{e, \vect{q}}[n]}_2^2 + \frac{4}{(\Delta r)^2} \, h_{\text{g}}^2 \, \abs{\zeta'(s[n \! - \! k_{\text{g}}])}^2 \norm{\vect{x}[n \! - \! k_{\text{g}}]}_2^2 \nonumber\\
	&\hspace{3em} \left(\vect{\nu}_{Q_{\text{sp}}}'[n \! - \! k_{\text{g}}]\tran \matx{B}_{Q_{\text{sp}}}^{\text{sp}} \vect{q}_{\text{I}, Q_{\text{sp}}, \iota[n - k_{\text{g}}]}[n \! - \! 1 \! - \! k_{\text{g}}]\right)^2 \\
	&b_{\mu,\text{Q}}[n] \nonumber\\
	&\hspace{0.5em} = \tau \norm{\tilde{\vect{g}}_{e, \vect{q}}[n]}_2^2 + \frac{4}{(\Delta r)^2} \, h_{\text{g}}^2 \, \abs{\zeta'(s[n \! - \! k_{\text{g}}])}^2 \norm{\vect{x}[n \! - \! k_{\text{g}}]}_2^2 \nonumber\\
	&\hspace{3em} \left(\vect{\nu}_{Q_{\text{sp}}}'[n \! - \! k_{\text{g}}]\tran \matx{B}_{Q_{\text{sp}}}^{\text{sp}} \vect{q}_{\text{Q}, Q_{\text{sp}}, \iota[n - k_{\text{g}}]}[n \! - \! 1 \! - \! k_{\text{g}}]\right)^2.
\end{align}
In the inequality for $e_{\text{I}}[n]$, we neglect $e_{\text{Q}}[n]$ and vice versa. The solutions of both inequalities are of the form ${0 \leq \mu[n] \leq 2/{b_{\mu,\text{I/Q}}}[n]}$. Since we did not include any coupling between $e_{\text{I}}[n]$ and $e_{\text{Q}}[n]$, the final upper bound for $\mu[n]$ is chosen conservatively as ${\mu_{\text{max}}[n] = 2/(b_{\mu,\text{I}}[n] + b_{\mu,\text{Q}}[n] + \xi)}$. This value is guaranteed to be smaller than or equal to $2/{b_{\mu,\text{I/Q}}[n]}$ since the $b_{\mu,\text{I/Q}}[n]$ are non-negative. Following this approach, we obtain the final step-size normalization by choosing
\begin{align}
	\mu[n] &= \mu \, \mu_{\text{max}}[n] = \mu \, \frac{2}{b_{\mu,\text{I}}[n] + b_{\mu,\text{Q}}[n] + \xi} \nonumber\\
	&= \mu \, \bigg(2 \frac{h_{\text{g}}^2}{(\Delta r)^2} \abs{\zeta'(s[n \! - \! k_{\text{g}}])}^2 \norm{\vect{x}[n \! - \! k_{\text{g}}]}_2^2\nonumber\\
	&\phantom{{} = \mu \, \bigg(} \abs{\vect{\nu}_{Q_{\text{sp}}}'[n \! - \! k_{\text{g}}]\tran \matx{B}_{Q_{\text{sp}}}^{\text{sp}} \vect{q}_{Q_{\text{sp}}, \iota[n - k_{\text{g}}]}[n \! - \! 1 \! - \! k_{\text{g}}]}^2 \nonumber\\
	&\phantom{{} = \mu \, \bigg(} + \tau \norm{\tilde{\vect{g}}_{e, \vect{q}}[n]}_2^2 + \xi \bigg)^{-1}
\end{align}
with the constant step-size ${0 \leq \mu \leq 1}$. In order to improve the adaptation rate and the performance consistency of the CIO\nobreakdash-WSAF, the \ac{TD} concept \eqref{equ:wsaf_td_concept} can be applied, too.

\subsection{Computational Complexity}

An important aspect in \ac{DSIM} applications is the computational complexity of the estimation algorithm, since it directly determines the power consumption and real-time capability. Therefore, we provide the general number of operations required by the CI-WSAF and the CIO-WSAF, depending on the filter length and the spline parameters. Additionally, we compare the complexity to two state-of-the-art algorithms for a specific configuration used in \secref{sec:perf_simulation}.

\vspace{5pt}
\subsubsection{CI-WSAF}
\begin{table*}[!t]
	\renewcommand{\arraystretch}{1.5}
	\caption{Arithmetic operations per sample of CI-WSAF and extensions.}
	\label{tab:complexity_ci_wsaf_general}
	\centering
	\scriptsize
	\begin{tabular}{ll|cccc}
		\hline
		\multicolumn{2}{l|}{Module} & Add./Sub. & Mult. & Div. & Sqrt. \\
		\hline
		\multicolumn{2}{l|}{CI-WSAF, \ $\zeta(s) = \abs{s}^2$} & $\begin{array}{l} 2 \, Q_{\text{lin}} \, (Q_{\text{out}}+3) + 2 \, Q_{\text{sp}}^2 \\{} + Q_{\text{sp}} \, (Q_{\text{out}}-2) + Q_{\text{out}} + N_{\text{sp}} - 1\end{array}$ & $\begin{array}{l} 2 \, Q_{\text{lin}} \, (Q_{\text{out}}+5) + 2 \, Q_{\text{sp}}^2 \\{} + Q_{\text{sp}} \, (Q_{\text{out}}+1) + Q_{\text{out}} + N_{\text{sp}} + 4 \end{array}$ & 0 & 0 \\
		\cline{3-6}
		\multicolumn{2}{l|}{Additional operations for $\abs{s}$} & 0 & 1 & 2 & 1\\
		\multicolumn{2}{l|}{Normalization} & $2 \, Q_{\text{lin}} + N_{\text{sp}}$ & $2 \, Q_{\text{lin}} + N_{\text{sp}} + 2$ & 1 & 0 \\
		\hline
		\multirowcell{2}[-1pt][l]{Constr. SGD} & $\ell_1$ norm & $4 \, Q_{\text{lin}}$ & $6 \, Q_{\text{lin}} - 1$ & $2 \, Q_{\text{lin}}$ & $Q_{\text{lin}}$ \\
		& $\ell_2$ norm & $4 \, Q_{\text{lin}}$ & $6 \, Q_{\text{lin}} - 1$ & 0 & 0\\
		\cline{3-6}
		\multirowcell{2}[0pt][l]{Norm limiter} & $\ell_1$  norm & $2 \, Q_{\text{lin}}$ & $2 \, Q_{\text{lin}}$ & 0 & $Q_{\text{lin}}$ \\
		& $\ell_2$ norm & $2 \, Q_{\text{lin}}$ & $2 \, Q_{\text{lin}}$ & 0 & 0\\
		\cline{3-6}
		\multicolumn{2}{l|}{Transf. input} & $4 \, Q_{\text{lin}} + 10$ & $6 \, Q_{\text{lin}} - 4$ & 0 & 0 \\
		\hline
		\multirowcell{2}[0pt][l]{1-tap scaler} & N-LMS & 3 & 4 & 1 & 0 \\
		& Weighted LS & 3 & 4 & 1 & 0 \\
		\hline
	\end{tabular}
\end{table*}%
In \tabref{tab:complexity_ci_wsaf_general}, we break down the output and update equations of the CI-WSAF and its extension into real-valued operations per input sample. Besides additions/subtractions and multiplications we separately list division and square root operations. Due to their complexity, the latter two are usually approximated. We do not detail possible division and square root implementations in this work. For example, in \cite{Flynn1997}, a low-complex approximation of the reciprocal is shown based on look-up tables. This method would imply an additional multiplication for every division, where the numerator is not 1. All operations that only involve constants are assumed to be precomputed. In \tabref{tab:complexity_ci_wsaf_general}, we used the fixed nonlinearity ${\zeta(s) = |s|^2}$ as a baseline, the additional operations required for ${\zeta(s) = |s|}$ are listed separately. Observing the spline basis matrices in \eqref{equ:b_spline_basis_mat_1_2}--\eqref{equ:cr_spline_basis_mat} reveals that the complexity of the product ${\vect{\nu}\tran \matx{B}^{\text{sp}} \vect{q}}$ heavily depends on the spline type. Besides the omission of products with zero, all powers of two would result in simple shift operations. However, in the following we assume a general $\matx{B}^{\text{sp}}$ to ensure generality of the results. We also consider a general output filter of length $Q_{\text{out}}$. Consequently, the values in \tabref{tab:complexity_ci_wsaf_general} represent an upper bound at the base sampling rate. If the \ac{Tx} \ac{BB} allocation necessitates an oversampling factor larger than 1, the effective complexity per \ac{Rx} sample at the \ac{Rx} \ac{BB} rate increases accordingly. In case of the norm limiter, we neglect the scaling of the filter weights if the norm exceeds the target value $\rho_w$, since this step occurs very seldomly. The \ac{DCT} (type-II) used by the \ac{TD} extension is optimized for delay-line inputs, thereby reducing its complexity significantly \cite{Kober2004}. This variant is also known as \ac{SCT}.

\vspace{5pt}
\subsubsection{CIO-WSAF}
\begin{table*}[!t]
	\renewcommand{\arraystretch}{1.5}
	\caption{Arithmetic operations per sample of CIO-WSAF (for extensions see \tabref{tab:complexity_ci_wsaf_general}).}
	\label{tab:complexity_cio_wsaf_general}
	\centering
	\scriptsize
	\begin{tabular}{ll|cccc}
		\hline
		\multicolumn{2}{l|}{Module} & Add./Sub. & Mult. & Div. & Sqrt. \\
		\hline
		\multicolumn{2}{l|}{CIO-WSAF, \ $\zeta(s) = \abs{s}^2$} & $\begin{array}{l} 2 \, Q_{\text{lin}} \, (Q_{\text{out}}+3) + 2 \, Q_{\text{sp}}^2 \\{} + Q_{\text{sp}} \, Q_{\text{out}} + 2 \, Q_{\text{out}} + 2 \, N_{\text{sp}} - 2 \end{array}$ & $\begin{array}{l} 2 \, Q_{\text{lin}} \, (Q_{\text{out}}+5) + 2 \, Q_{\text{sp}}^2 \\{} + Q_{\text{sp}} \, (Q_{\text{out}}+3) + 2 \, Q_{\text{out}} + 2 \, N_{\text{sp}} + 6\end{array}$ & 0 & 0 \\
		\cline{3-6}
		\multicolumn{2}{l|}{Additional operations for $\abs{s}$} & 0 & 1 & 1 & 1 \\
		\multicolumn{2}{l|}{Normalization} & $2 \, Q_{\text{lin}} + 2 \, N_{\text{sp}}$ & $2 \, Q_{\text{lin}} + 2 \, N_{\text{sp}} + 2$ & 1 & 0 \\
		\hline
	\end{tabular}
\end{table*}%
The baseline complexity of the CIO-WSAF for ${\zeta(s) = |s|^2}$ is given in \tabref{tab:complexity_cio_wsaf_general}. There, the additional operations for ${\zeta(s) = |s|}$ and the step-size normalization are provided, too. The extensions for controlling the norm of the weight vector and for decorrelating the input data are unaffected by the complex control points. Thus, their complexity can be found in \tabref{tab:complexity_ci_wsaf_general}. Since the CIO-WSAF provides a complex-valued output, the single-tap scaler can be omitted. Overall, the costs of the CI-WSAF and CIO-WSAF are comparable, especially for short output filters.

\vspace{5pt}
\subsubsection{Comparison to State-of-the-Art Concepts}
The first chosen state-of-the-art algorithm is the IM2LMS \cite{Gebhard2017}, \iac{LMS} variant, which is limited to the cancellation of second-order \ac{IMD}. It requires the Q-path interference signal to be derived from the I-path by scaling. Thus, it is combined with the single-tap \ac{LS} scaler \eqref{equ:wsaf_1_tap_weighted_ls}. The second method we use for comparison is the \ac{KRLS} algorithm, a very general adaptive learning concept \cite{Engel2004}. We choose a real-valued Gaussian kernel with a complex-valued input \cite{Boloix2019}, which allows the \ac{KRLS} to model a wide range of complex-valued nonlinearities without the single-tap scaler. In order to limit its complexity, the \ac{KRLS} needs an additional sparsification method, in our case the \ac{ALD} criterion. It maintains a growing dictionary $\mathcal{D}$ of relevant input vectors, combined with a set of complex weights. In stationary scenarios, the dictionary size $\abs{\mathcal{D}}$ can be considered to be settled at some point. In the \ac{IMD} cancellation scenarios in \secref{sec:perf_simulation}, the typical dictionary size is 125, with maximum values of up to 700 for high leakage powers.

\tabref{tab:complexity_imdx_algo_comp} explicitly gives the number of operations per sample for all discussed algorithms applied to \ac{IMD} interference cancellation. The length of the input vector $Q_{\text{lin}}$ is chosen to be 16 and $Q_{\text{sp}}$ is 3. The spline-based methods use a normalized step-size, the weight norm limiter with ${p = 1}$ and the TD extension. In addition, the output filter is a simple delay, which allows for significant computational savings. In the equations in \tabref{tab:complexity_ci_wsaf_general} this can be reflected by replacing $N_{\text{sp}}$ with $Q_{\text{sp}}$. Hence, both spline models are considered to be implementable for real-time cancellation. The higher complexity of the CI(O)-WSAF compared to the IM2LMS is clearly outweighed by its much higher flexibility. The numbers for the \ac{KRLS} assume a settled dictionary of given size, where only the weights are adjusted. The Gaussian kernel requires the evaluation of $\exp(x)$, which would be approximated in practice. Still, the \ac{KRLS} requires immense computing power for online cancellation at typical \ac{LTE}/\ac{NR} sampling rates. Even the sole prediction of new samples without altering the weights is too expensive for mobile devices. In the adaptation phase, the CIO-WSAF features less than \SI{0.6}{\percent} of the additions and multiplications of the \ac{KRLS}.
\begin{table}[!t]
	\renewcommand{\arraystretch}{1.5}
	\caption{Exemplary complexity of DSIM algorithms for IMD cancellation.}
	\label{tab:complexity_imdx_algo_comp}
	\centering
	\scriptsize
	\begin{tabular}{ll|rrrrr}
		\hline
		\multicolumn{2}{l|}{Algorithm} & Add./Sub. & Mult. & Div. & Sqrt. & Exp. \\
		\hline
		\multicolumn{2}{l|}{IM2LMS + 1-tap scaler} & $164$ & $201$ & $2$ & $0$ & $0$ \\
		\hline
		\multicolumn{2}{l|}{CI-WSAF + 1-tap scaler} & $290$ & $353$ & $2$ & $16$ & $0$ \\
		\multicolumn{2}{l|}{CIO-WSAF} & $299$ & $363$ & $1$ & $16$ & $0$ \\
		\hline
		\multirowcell{2}[0pt][l]{KRLS,\\$\abs{\mathcal{D}} = 125$} & Prediction & $8125$ & $4375$ & $0$ & $0$ & $125$ \\
		& Pred. + Train. & $7.1 \!\cdot\! 10^4$ & $6.7 \!\cdot\! 10^4$ & $125$ & $0$ & $125$ \\
		\multirowcell{2}[0pt][l]{KRLS,\\$\abs{\mathcal{D}} = 700$} & Prediction & $4.6 \!\cdot\! 10^4$ & $2.5 \!\cdot\! 10^4$ & $0$ & $0$ & $700$ \\
		& Pred. + Train. & $2.0 \!\cdot\! 10^6$ & $2.0 \!\cdot\! 10^6$ & $700$ & $0$ & $700$ \\
		\hline
	\end{tabular}
\end{table}%

\section{Performance Evaluation}
\label{sec:perf_simulation}

We conclude our investigations on spline-based \ac{DSIM} algorithms with performance simulations on a realistic \ac{IMD} interference scenario. Besides the steady-state cancellation with and without output delays, we also show the effect of the norm constraint on the filter weights.

\subsection{Setup}
\begin{figure}
	\centering
	\includegraphics[width=0.76\columnwidth]{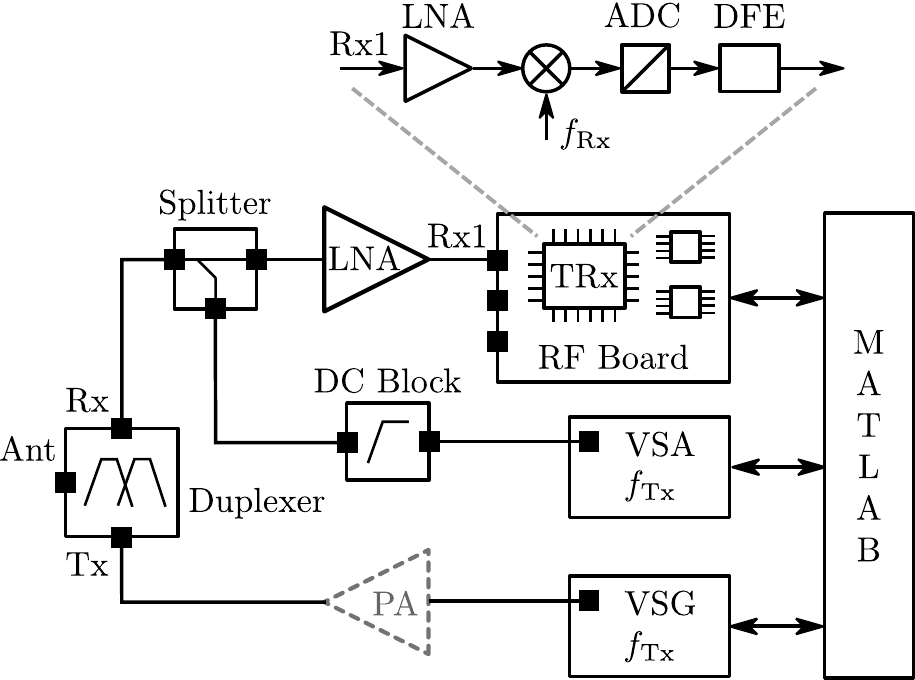}
	\caption{Measurement setup for characterization of duplexer and Rx nonlinearity. Unused ports are terminated.}
	\label{fig:Meas_Setup}
\end{figure}%
The interference model used in the following is based on measurements on an integrated \ac{CMOS} transceiver using the setup outlined in \figref{fig:Meas_Setup}. The \ac{Tx} signal is generated using \iac{VSG} and optionally amplified by a discrete \ac{PA}. A discrete duplexing filter provides the Tx-Rx isolation in the \ac{AFE}. The signal after the duplexer is captured by \iac{VSA} in order to obtain the actual input signal of the \ac{Rx} chain. Simultaneously, the signal is passed to an external \ac{LNA} with a gain of \SI{15}{\dB}. Unlike the internal \ac{LNA} of the subsequent \ac{RF} transceiver, the external \ac{LNA} is assumed to be sufficiently linear to not contribute to the \ac{IMD}. On the one hand, this setup enables a characterization of multiple duplexers for different \ac{LTE} bands. On the other hand, it allows to extract the \ac{IMD} components from the \ac{Rx} \ac{BB} sequence. The \ac{IMD} interference power was measured for various leakage power levels $P(y_{\text{BB}}^{\text{TxL}})$ at the chip input, leading to the nonlinearity coefficients $\gamma_k$ as used in \eqref{equ:imdx_mdl_bb}.
\begin{figure}
	\centering
	{\footnotesize
		\begin{tabular}{lr}
			(a) & \parbox{0.8\columnwidth}{\raggedleft\includegraphics[width=0.78\columnwidth]{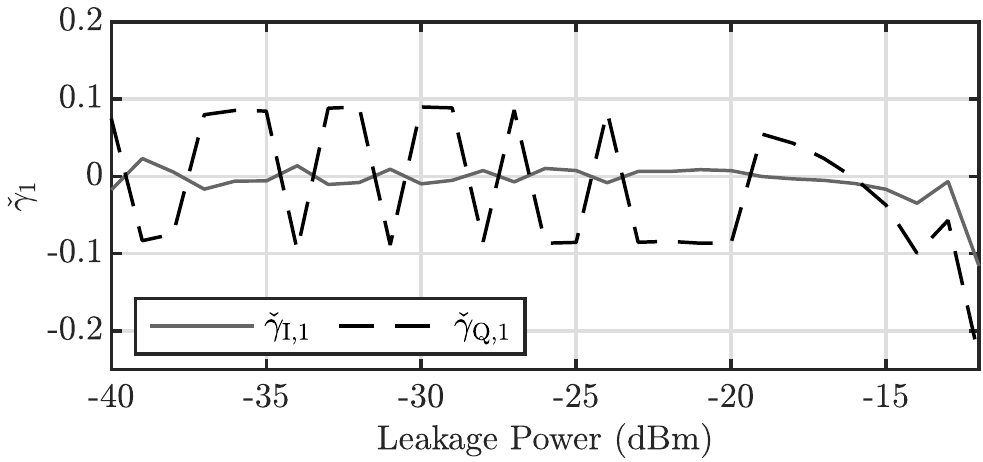}}\\
			&\\
			(b) & \parbox{0.8\columnwidth}{\raggedleft\includegraphics[width=0.76\columnwidth]{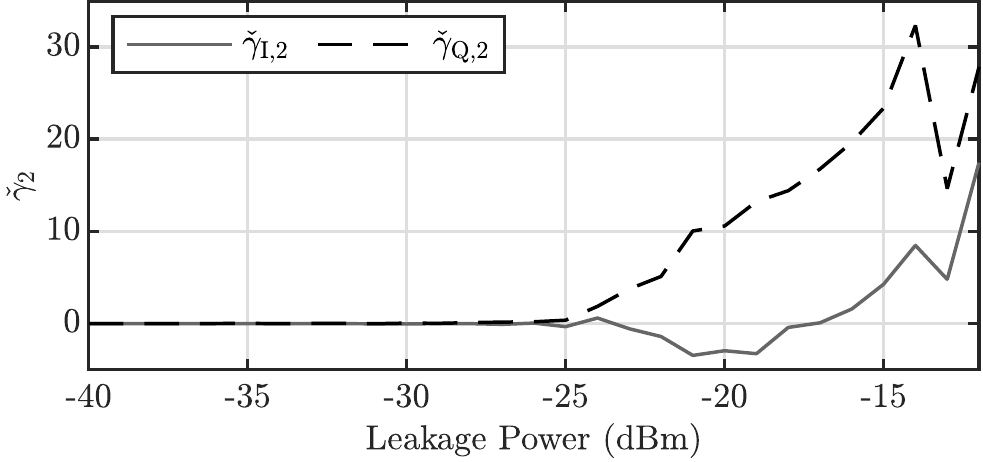}}\\
			&\\
			(c) & \parbox{0.8\columnwidth}{\raggedleft\includegraphics[width=0.8\columnwidth]{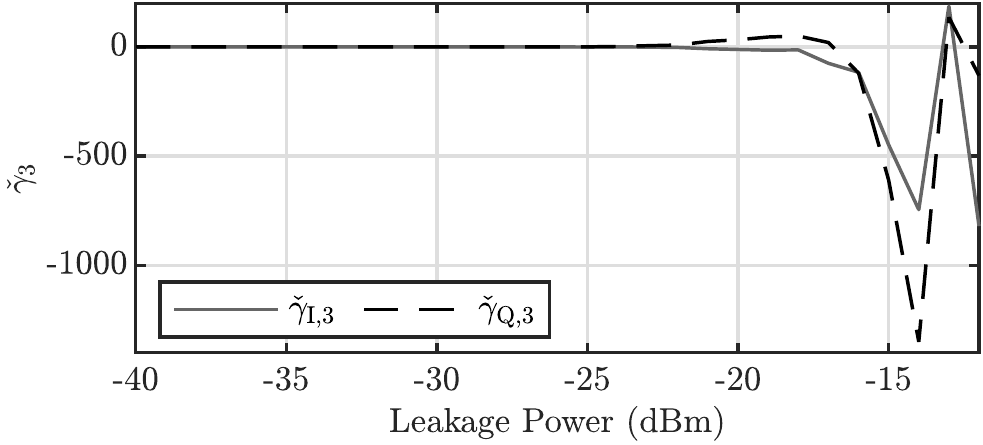}}\\
		\end{tabular}
	}
	\caption{Input-referred coefficients of intermodulation products in \ac{RF} receiver: (a) IMD2, (b) IMD4 and (c) IMD6.}
	\label{fig:IMD_coeffs}
\end{figure}%
\figref{fig:IMD_coeffs} depicts the real and imaginary parts of the input-referred coefficients ${\check{\gamma}_k = \gamma_k / A_{\text{lin}}}$, which differ considerably. The values are not constant, since for low input power levels, the \ac{IMD} products are below the noise floor, thereby increasing the error of the fit. Additionally, for high input powers, compression effects occur that are not covered by the \ac{IMD} modeling provided in \secref{sec:imd_self_intf}. 

In addition to the receiver nonlinearity, we take the \ac{Tx} distortions caused by the \ac{PA} into account. This is achieved by replacing the ideal \ac{Tx} \ac{RF} model \eqref{equ:tx_rf_mdl_ideal} by
\begin{equation}
	x_{\text{RF}}(t) = \Re\left\{g_{\text{BB}}^{\text{PA}}(x_{\text{BB}}(t)) \, e^{\imag 2\pi f_{\text{Tx}} t}\right\}
\end{equation}
with the equivalent \ac{BB} transfer characteristic $g_{\text{BB}}^{\text{PA}}(.)$ of the \ac{PA}. With this model, we assume that the power of any harmonic emissions caused by the \ac{PA} is much lower than the in-band signal power. For $g_{\text{BB}}^{\text{PA}}(.)$, we choose the well-known Rapp model, which is a memoryless behavioral model for solid-state amplifiers \cite{Glock2015, Rapp1991}:
\begin{equation}
	g_{\text{BB}}^{\text{PA}}(x_{\text{BB}}(t)) = \frac{A_{\text{PA}} \abs{x_{\text{BB}}(t)}}{\left(1 + \left(\frac{A_{\text{PA}} \abs{x_{\text{BB}}(t)}}{x_{\text{max}}}\right)^{2p}\right)^{\frac{1}{2p}}} \, e^{\imag \arg(x_{\text{BB}}(t))}.
\end{equation}
This model describes the nonlinear AM-AM\footnote{Amplitude modulation} conversion by means of a smooth saturation characteristic, while assuming that the AM-PM\footnote{Phase modulation} conversion is neglible. The smoothness of the transition from the linear part to the saturation region is given by the parameter $p$, the output saturation level is specified by $x_{\text{max}}$. In the following simulations, $A_{\text{PA}}$ is varied to reflect different interference power levels. Thus, instead of using an absolute value for $x_{\text{max}}$, we define it implicitly by means of the \ac{CR}
\begin{equation}
	\text{CR} = 20 \log_{10} \frac{x_{\text{max}}}{A_{\text{PA}} \sigma_x}.
\end{equation}
$\sigma_x$ is the \ac{RMS} value of $x_{\text{BB}}(t)$. We approximate the behavior of the discrete \ac{PA} in our measurement setup with the parameters ${p = 2}$ and ${\text{CR} = \SI{6}{\dB}}$.

The leakage path $\tilde{h}_{\text{BB}}^{\text{TxL}}[n]$ is modeled by means of fitted \ac{FIR} impulse responses, which are based on the measured duplexer stop-band frequency responses. All impulse responses contain 21 values, which decay towards higher delays. The Tx-Rx isolation is about \SI{50}{\dB}.

The \ac{Tx} signal is an LTE\nobreakdash-20 \ac{UL} signal with a 16-QAM\footnote{Quadrature Amplitude Modulation} alphabet and 10 \acp{RB} allocated in the index range $[10, 19]$, where the index 1 is at the lower end of the \ac{BB} spectrum. The \ac{Rx} signal is chosen to be a fully allocated LTE\nobreakdash-20 \ac{DL} signal, which corresponds to a utilized bandwidth of \SI{18}{\MHz}. Note that LTE\nobreakdash-20 is used due to limitations of the \ac{RF} setup. Clearly, the proposed algorithms are applicable to the higher bandwidths defined by the 5G standard, too. The power of the \ac{Rx} signal at the chip input is \SI{-72}{\dBm}, which is close to reference sensitivity \cite{3gpp.36.211_14} at the input of the external \ac{LNA}. The \ac{SNR} without \ac{IMD} is about \SI{10}{\dB}.

In all scenarios, the \ac{DSIM} is applied before the \ac{CSF}. For a successful interference cancellation, a correct time-alignment between the \ac{Tx} and the total \ac{Rx} signal has to be ensured. In an online operation, this could be achieved by applying a correlation-based adaptive synchronization on the strongest \ac{IMD} component \cite{Paireder2019} or by utilizing a known constant delay. Using the described parameters, exemplary signals are generated in a simulation based on \eqref{equ:imdx_mdl_bb} and the average cancellation performance is computed. This evaluation approach is vital especially for \ac{LMS}-type algorithms, since they typically suffer from a high performance spread. In the simulation, we neglect any intermodulation products between the leakage signal, the wanted \ac{Rx} signal and noise due to the wide bandwidth and low power of the wanted and noise components. Even if this assumption was not fulfilled, our simplification would only affect the optimum \ac{SNR}, but not the relative performance between several \ac{DSIM} algorithms. Due to the narrow allocation, no oversampling is required for the analyzed \ac{IMD} products, \ie up to the sixth order.

\subsection{Scaled Nonlinearity in Q-Path}

In the first test case, we assume a scalar coupling between the nonlinearities in the I- and Q-path. While this assumption is not fulfilled by the targeted receiver, it enables a comparison between the CI\nobreakdash-WSAF and the comparably low-complex IM2LMS. In addition, we include the computationally intensive, but very general, \ac{KRLS} algorithm. Since the ratio between the measured coefficients $\gamma_{k,\text{I}}$ and $\gamma_{k,\text{Q}}$ changes substantially over the leakage power range, we alter these values for this simulation. We keep the interference power unchanged and assume a coupling of ${\delta_{\text{Q}} = -1}$, leading to the modified coefficients
\begin{equation}
	\tilde{\gamma}_k = (1 - \imag) \, \frac{\sign(\gamma_{k, \text{I}}) \abs{\gamma_k}}{\sqrt{2}}.
\end{equation}
The value of $\delta_{\text{Q}}$ does not influence the performance, since it is reliably estimated by the single-tap scaler.

The cancellation performance of the algorithms is compared by means of the \ac{SINR}
\begin{equation}
	\text{SINR} \coloneqq \frac{\expv\left[\abs{\check{y}_{\text{BB}}^{\text{Rx}}[n]}^2 \right]}{\expv\left[\abs{y_{\text{BB}}^{\text{IMD}}[n] - \hat{y}_{\text{BB}}^{\text{IMD}}[n] + \check{\eta}_{\text{BB}}[n]}^2 \right]},
\end{equation}
which is averaged over two \ac{LTE} slots, while excluding the first symbol. Thus, the initial convergence does not affect the measured performance.
\begin{figure}
	\centering
	\includegraphics[width=0.9\columnwidth]{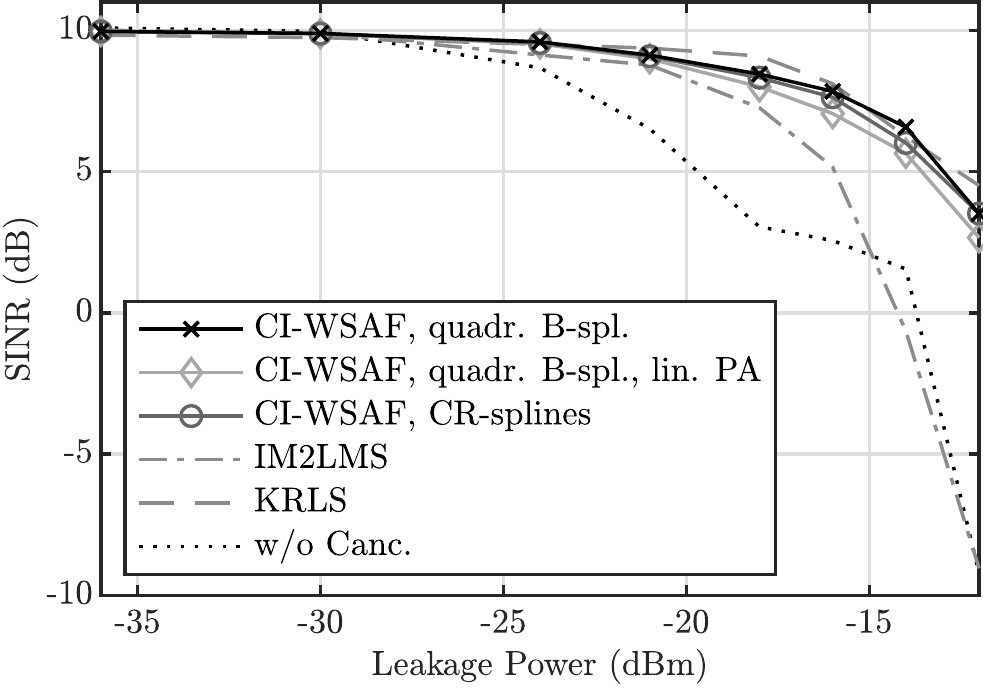}
	\caption{Steady-state SINR improvement for various IMD cancellation schemes in case of scaled nonlinearity in Q-path.}
	\label{fig:SINR_scaledQ}
\end{figure}%
\figref{fig:SINR_scaledQ} depicts the \ac{SINR} for two variants of the CI\nobreakdash-WSAF and two algorithms for comparison. In addition, we show the impact of the \ac{PA} model. Each value is the ensemble average over the results for six different fitted duplexer impulse responses, where for each duplexer 50 runs with randomly generated signals were performed. The linear filter length of all algorithms is set to 16. All relevant parameters are optimized for the individual leakage power levels to guarantee the highest possible performance for a fair comparison. In case of the IM2LMS, previous knowledge about the sign of the real part of the IMD2 component is required in order to ensure correct operation. In practice, this information could be obtained by correlating $\abs{x_{\text{BB}}[n]}^2$ with $y_{\text{BB}}^{\text{Tot}}[n]$. Since the output of the IM2LMS is real-valued, the Q-path interference is estimated using \eqref{equ:wsaf_1tap_q_scaling}, similar to the CI\nobreakdash-WSAF. Because of its superior performance, we employ the weighted \ac{LS} solution with ${\lambda_{\text{cpl}} = 0.9998}$ for this purpose. The \ac{DC} cancellation in the receiver is replicated by applying the notch filter ${H(z) = (1 \! - \! z^{-1})/(1 \! - \! 0.998 z^{-1})}$ to the IM2LMS output. The step-size of the IM2LMS is selected in the range $(0, 0.056]$, where extremely small values are used for low interference powers. The regularization parameter is chosen within $[0.01, 1]$. Despite the necessity of prior knowledge, the IM2LMS shows the lowest \ac{SINR} among the compared algorithms. The CI\nobreakdash-WSAF is used in the \ac{TD} variant \eqref{equ:wsaf_td_concept} with the norm limiting \eqref{equ:wsaf_norm_limiter}. In order to improve the adaptation rate of the spline-based algorithm, the weights of the linear section are initialized to random constants, which are unaltered for all runs. The fixed nonlinearity is ${\zeta(s) = |s|^2}$ to avoid a square root in the feedback loop.
\begin{table}[!t]
	\renewcommand{\arraystretch}{1.5}
	\caption{Parameter values of CI-WSAF for IMD cancellation.}
	\label{tab:sim_ci_wsaf_params}
	\centering
	\begin{tabular}{l|l||l|l}
	\hline
	Parameter & Value(s) & Parameter & Value(s)\\
	\hline
	$Q_{\text{lin}}$ & $16$ & $\tau$ & $[1, 660]$\\
	$\vect{w}[0]$ & $\sim \mathcal{N}_{Q_{\text{lin}}}(\vect{0}, 0.01 \, \matx{I})$ & $N_\text{sp}$ & $20$ \\
	$p$ & 1 & $Q_{\text{sp}}$ & $3$, $4$ \\
	$\rho_w$ & 3 & $r_0$ & $-0.1$, $-0.15$ \\
	$\mu$ & $[0.0015, 0.015]$ & $\Delta r$ & $0.05$ \\
	$\xi$ & $[0.0023, 0.45]$ & & \\
	\hline
	\end{tabular}
\end{table}%
\tabref{tab:sim_ci_wsaf_params} summarizes the chosen parameter ranges for the CI\nobreakdash-WSAF, where most values tend to become larger with increasing leakage power. $\mathcal{N}_{Q_{\text{lin}}}$ denotes a multivariate normal distribution. The parameter $\lambda_{\text{cpl}}$ of the single-tap scaler was again $0.9998$. A negative value for the first knot $r_0$ enables a domain of the spline function down to 0, as required by the fixed nonlinearity $|s|^2$. We have ${h_{\text{out}}[n] = 1}$, since we do not employ any pipelining and the cancellation is performed before the \ac{CSF}. In \figref{fig:SINR_scaledQ}, the CI\nobreakdash-WSAF shows excellent performance, clearly outperforming the IM2LMS by up to \SI{12.5}{\dB}. For all leakage powers levels, it restores the \ac{SINR} to values above the \SI{0}{\dB} threshold. For the same number of control points, there is no benefit of using cubic CR-splines instead of quadratic B-splines, likely due to the smoothness of the \ac{IMD} nonlinearity. Compared to a linear \ac{PA} model, the clipping of the signals peaks by the Rapp \ac{PA} model slightly improves the performance of the WSAF algorithms. Without clipping, signal peaks occur very seldomly, leading to slow adaptation of the corresponding spline control points and, in further consequence, higher estimation errors. Another minor \ac{SINR} improvement is achievable by using an \ac{LS}-based learning algorithm, in our case the \ac{KRLS}. For this simulation, we used a complexified Gaussian kernel with a standard deviation in the range ${[5, 100]}$. The dictionary size was limited by means of the \ac{ALD} approach with a threshold in the range ${[10^{-5},5 \cdot 10^{-5}]}$. Due to its fast convergence and high complexity, the \ac{KRLS} weights were adapted for the first three \ac{LTE} symbols, or about 6600 samples, only. Its maximum \ac{SINR} advantage over the CI\nobreakdash-WSAF is about \SI{1}{\dB}. A practical aspect to improve the convergence of all algorithms is to disable the adaption for all samples close to the symbol boundaries within $x_{\text{BB}}[n]$. This measure avoids high error values, which would be caused by the bandwidth increase at symbol transitions \cite{Motz2020_C2}.

\begin{figure}
	\centering
	\includegraphics[width=0.9\columnwidth]{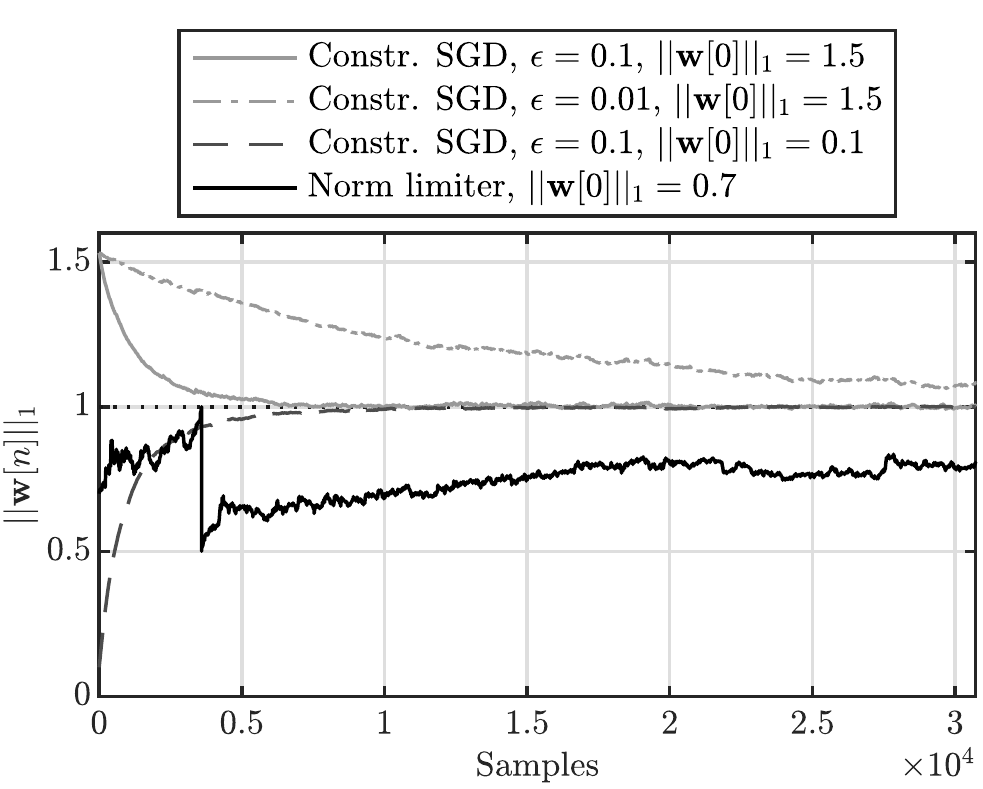}
	\caption{Adaptation of $\ell_1$ norm of filter weights due to constrained SGD optimization and limiter.}
	\label{fig:Constr_SGD_Comp}
\end{figure}%
Another interesting aspect of the CI-WSAF is the constraint on the filter weights, which helps to avoid internal clipping. In \figref{fig:Constr_SGD_Comp}, the evolution of the $\ell_1$ norm of $\vect{w}[n]$ is compared for a leakage power of \SI{-18}{\dBm} and different parameters of the constraint. Using the constrained \ac{SGD} method with a weighting of ${\epsilon = 0.1}$, $\norm{\vect{w}[n]}_1$ convergences close to the desired value of 1 within three symbols. A weighting of ${\epsilon = 0.01}$ already slows down the adaptation substantially, but still the norm tends towards 1, thereby reducing the clipping. Although this method shows the expected results, the second optimization goal potentially impacts the steady-state interference cancellation by slowing down the overall adaptation. In contrast, the norm limiting is also able to reduce clipping while not impacting the steady-state performance at all. However, the weight rescaling in the initial phase leads to a performance disadvantage in the first 5000 samples.

\subsection{Independent Nonlinearity in I- and Q-Path}

In the second major test case, we allow for independent coefficients $\gamma_{k,\text{I}}$ and $\gamma_{k,\text{Q}}$, a requirement indicated by measurements on the test receiver. The IM2LMS and the CI-WSAF are not suitable for this scenario, thus, we resort to a comparison between the CIO\nobreakdash-WSAF and the \ac{KRLS}.
\begin{figure}
	\centering
	\includegraphics[width=0.85\columnwidth]{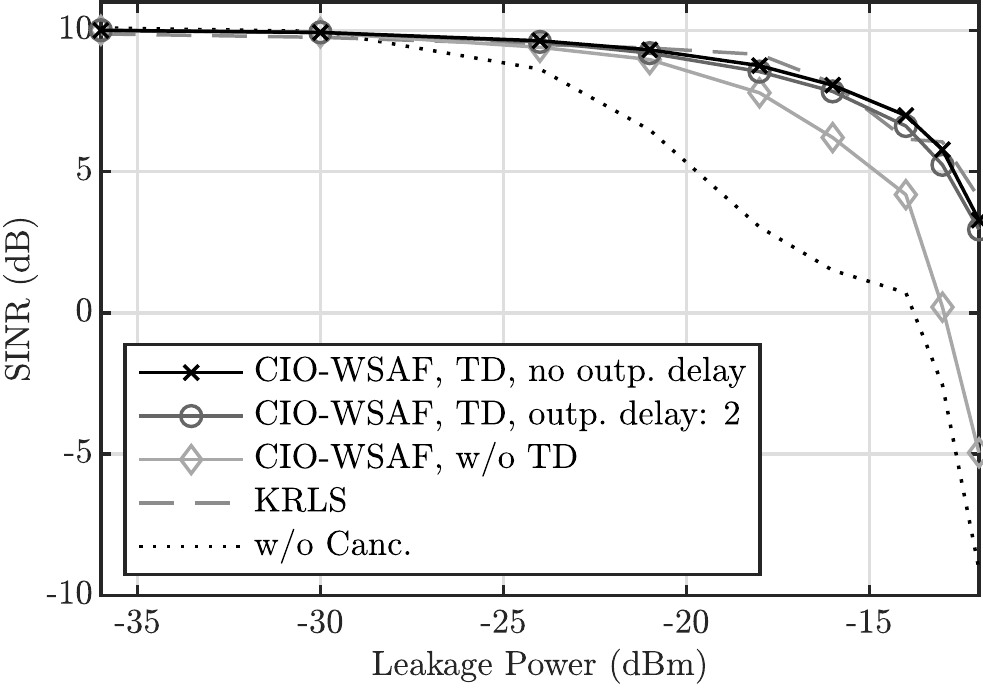}
	\caption{Steady-state SINR improvement for general IMD problem with independent nonlinearities in I- and Q-path.}
	\label{fig:SINR_indepIQ}
\end{figure}%
The first metric we analyze is again the \ac{SINR}, depicted in \figref{fig:SINR_indepIQ}. The overall \ac{SINR} degradation without any countermeasures is similar to the previous section, but now the underlying number of parameters is higher. Interestingly, this causes the previous small performance advantage of the \ac{KRLS} to vanish completely. Again, its kernel standard deviation was chosen in the range $[4.8, 141]$, whereas the \ac{ALD} threshold was chosen within $[10^{-5},5 \cdot 10^{-5}]$. The \ac{KRLS} weights and the dictionary were adapted for the first three \ac{LTE} symbols. Unlike the \ac{KRLS}, the CIO-WSAF with transformed input exhibits a slight \ac{SINR} improvement compared to the CI-WSAF. The simulation shows that the approximated step-size normalization of the CIO-WSAF behaves just as the standard version used in the CI-WSAF. Compared to \tabref{tab:sim_ci_wsaf_params}, the step-size ${\mu \in [4.4 \cdot 10^{-4}, 0.025]}$, the coupling factor ${\tau \in [1.3, 1200]}$ and the regularization ${\xi \in [0.017, 0.16]}$ required readjustments. The other parameters could be reused. Since no output filter is present, we have ${h_{\text{g}} = 1}$ and ${k_{\text{g}} = 0}$. We only consider quadratic interpolation, thus, $Q_{\text{sp}} = 3$. In \figref{fig:SINR_indepIQ}, we included two other variants of the CIO\nobreakdash-WSAF. One incorporates a delay of two samples in the output signal, possibly caused by pipelining stages in the design. In the algorithm, this configuration requires ${k_{\text{g}} = 2}$. The performance cost of this measure is minor, supporting a real-time implementation of our approach. The second WSAF variant we included does not use a transformed input vector for the linear section and has no output delay (\ie ${k_{\text{g}} = 0}$). Due to the properties of the \ac{LTE} \ac{UL} signal, which acts as a reference, the omission of the approximate decorrelation leads to \iac{SINR} drop of up to \SI{8.2}{\dB}.

\begin{figure}
	\centering
	\includegraphics[width=0.86\columnwidth]{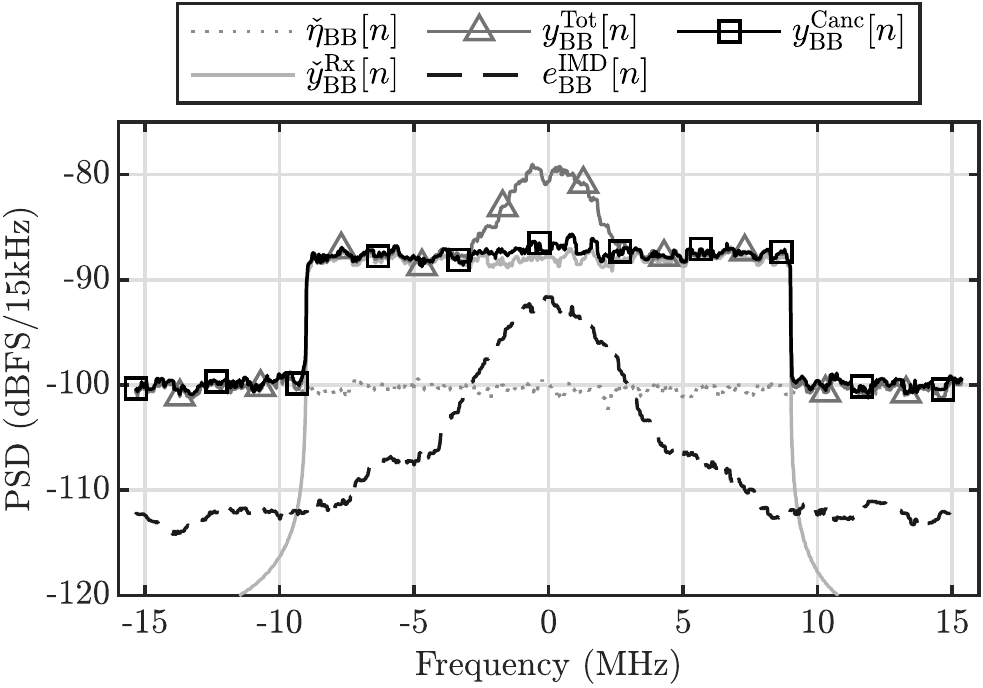}
	\caption{PSD of important Rx signal components before and after IMD cancellation using CIO-WSAF.}
	\label{fig:PSD_indepIQ}
\end{figure}%
In addition to the integral performance quantified by the \ac{SINR}, the \ac{PSD} of the \ac{Rx} signal components in \figref{fig:PSD_indepIQ} gives a descriptive visualization of the cancellation process. As a baseline, we show the noise floor $\check{\eta}_{\text{BB}}[n]$ and the ideal \ac{Rx} signal $\check{y}_{\text{BB}}^{\text{Rx}}[n]$, which, combined with the \ac{IMD} interference, form the total \ac{Rx} signal $y_{\text{BB}}^{\text{Tot}}[n]$. In this example, we chose a leakage power of \SI{-14}{\dBm}. For interference cancellation, we apply the CIO-WSAF with transformed input and no pipelining, which generates the interference replica $\hat{y}_{\text{BB}}^{\text{IMD}}[n]$. After \ac{DSIM}, the remaining interference ${e_{\text{BB}}^{\text{IMD}}[n] = y_{\text{BB}}^{\text{IMD}}[n] - \hat{y}_{\text{BB}}^{\text{IMD}}[n]}$ is below the \ac{Rx} level, which acts as additional noise from the perspective of the CIO-WSAF. The signal after \ac{DSIM}, $y_{\text{BB}}^{\text{Canc}}[n]$, closely resembles the wanted \ac{Rx} signal. This indicates a successful cancellation, which agrees with the \ac{SINR} of \SI{7}{\dB} shown in \figref{fig:SINR_indepIQ}.

\begin{figure}
	\centering
	\includegraphics[width=0.85\columnwidth]{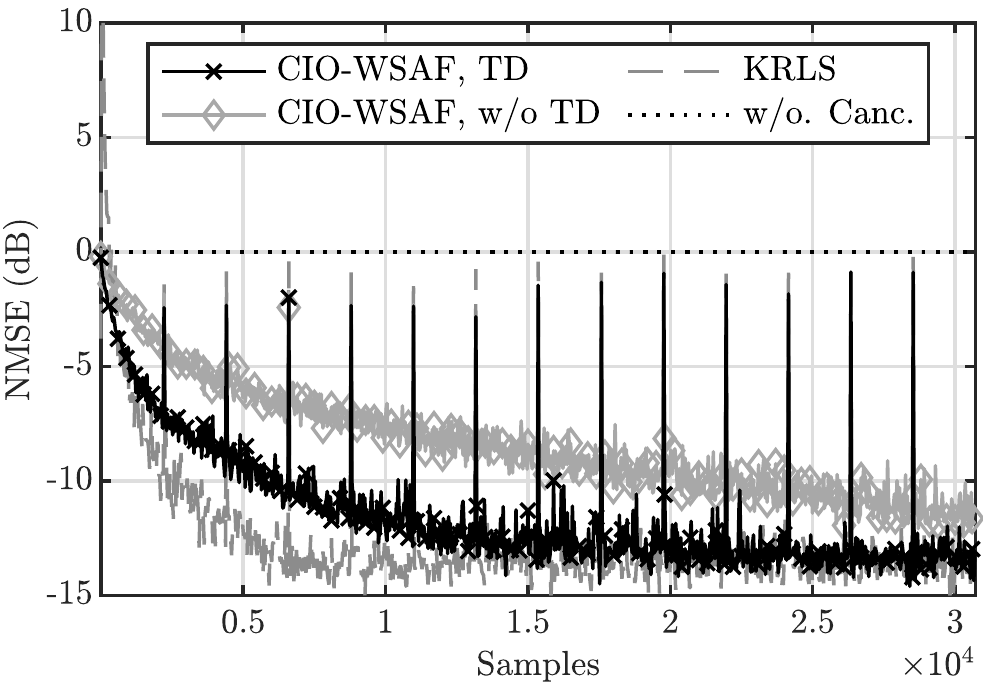}
	\caption{Adaptation behavior of IMD cancellation schemes in case of independent nonlinearities in I- and Q-path.}
	\label{fig:NMSE_indepIQ}
\end{figure}%
Another important metric for \ac{DSIM} applications is the adaptation time of the estimation algorithm. Therefore, in \figref{fig:NMSE_indepIQ} we compare the \ac{NMSE}
\begin{equation}
	\text{NMSE}[n] \coloneqq \frac{\expv\left[ \abs{y_\text{BB}^\text{IMD}[n] - \hat{y}_\text{BB}^\text{IMD}[n]}^2 \right]}{\expv\left[ \abs{y_\text{BB}^\text{IMD}[n]}^2 \right]}
\end{equation}
for two \ac{SAF} variants and the \ac{KRLS} at a leakage power level of \SI{-18}{\dBm}. As expected for RLS-type algorithms, the \ac{KRLS} features very fast adaptation within the training sequence of $6600$ samples. In contrast, the CIO\nobreakdash-WSAF without an input transform requires more than $3 \cdot 10^4$ samples to reach steady-state. The \ac{TD} variant manages to reduce the adaptation time to about ${2 \cdot 10^4}$ samples, which is a remarkable improvement for an \ac{SGD} algorithm.

\section{Conclusion}

We presented two novel adaptive learning schemes based on spline interpolation that allow for low-complex digital cancellation of transceiver self-interference caused by higher-order \ac{IMD}. Based on a comprehensive modeling of the receiver nonlinearities, we derived the \ac{BB} interference model, allowing to extract a Wiener structure suitable for online estimation. We proposed several extensions to the spline-based algorithm and precisely assessed the computational complexity of all modules. The interference cancellation performance was evaluated in practically relevant \ac{IMD} scenarios, which were based on measured receiver parameters. Compared to a general nonlinear estimation approach, our algorithms showed similar accuracy at a fraction of the computational costs.

\ifCLASSOPTIONcaptionsoff
  \newpage
\fi

\bibliographystyle{IEEEtran}
\bibliography{IEEEabrv,biblio}

\end{document}